\documentclass{elsarticle}

\usepackage[reviewcopy]{adndt}%{adndt} at NSCL, {../adndt} at Gross
\usepackage{longtable}

\usepackage{amsmath}

\usepackage{amssymb}

\biboptions{square,sort&compress}
\bibpunct[]{[}{]}{,}{n}{}{;}
\begin{document}

\begin{frontmatter}

\journal{Atomic Data and Nuclear Data Tables}

%% Author, fill in article title here

\title{Discovery of the astatine, radon, francium, and radium isotopes}

%% Fill in author list here
\author{C. Fry}
\author{M. Thoennessen\corref{cor1}}\ead{thoennessen@nscl.msu.edu}

 \cortext[cor1]{Corresponding author.}

 \address{National Superconducting Cyclotron Laboratory and \\ Department of Physics and Astronomy, Michigan State University, \\ East Lansing, MI 48824, USA}

\begin{abstract}
Currently, thirty-nine astatine, thirty-nine radon, thirty-five francium, and thirty-four radium isotopes have so far been observed; the discovery of these isotopes is discussed. For each isotope a brief summary of the first refereed publication, including the production and identification method, is presented.
\end{abstract}

\end{frontmatter}

%%% Keywords and subject classification are not used in ADNDT
%%%\begin{keywords}
%%%Insert list of keywords here.
%%%\end{keywords}

%%%\begin{subject}[Insert header for classifications]
%%%Use only if your journal has a subject classification requirement
%%%\end{subject}

%%% The table of contents should start a new page. This command will
%%% automatically pull all the unstarred \section, \subsection and
%%% \subsection titles into the Contents. Starred versions need to be
%%% done manually using
%%%            \addcontentsline{toc}{[[sub]sub]section}{Section title}
%%% at the correct place. Examples are given below.

%%% The lists of data figures and data tables are created automatically
%%% by the \listofDfigures and \listofDtables commands.

\newpage
\tableofcontents
%%\listofDfigures
\listofDtables

\vskip5pc

\section{Introduction}\label{s:intro}

The discovery of astatine, radon, francium, and radium isotopes is discussed as part of the series summarizing the discovery of isotopes, beginning with the cerium isotopes in 2009 \cite{2009Gin01}. Guidelines for assigning credit for discovery are (1) clear identification, either through decay-curves and relationships to other known isotopes, particle or $\gamma$-ray spectra, or unique mass and Z-identification, and (2) publication of the discovery in a refereed journal. The authors and year of the first publication, the laboratory where the isotopes were produced as well as the production and identification methods are discussed. When appropriate, references to conference proceedings, internal reports, and theses are included. When a discovery includes a half-life measurement the measured value is compared to the currently adopted value taken from the NUBASE evaluation \cite{2003Aud01} which is based on the ENSDF database \cite{2008ENS01}. In cases where the reported half-life differed significantly from the adopted half-life (up to approximately a factor of two), we searched the subsequent literature for indications that the measurement was erroneous. If that was not the case we credited the authors with the discovery in spite of the inaccurate half-life. All reported half-lives inconsistent with the presently adopted half-life for the ground state were compared to isomer half-lives and accepted as discoveries if appropriate following the criterium described above.

The first criterium is not clear cut and in many instances debatable. Within the scope of the present project it is not possible to scrutinize each paper for the accuracy of the experimental data as is done for the discovery of elements \cite{1991IUP01}. In some cases an initial tentative assignment is not specifically confirmed in later papers and the first assignment is tacitly accepted by the community. The readers are encouraged to contact the authors if they disagree with an assignment because they are aware of an earlier paper or if they found evidence that the data of the chosen paper were incorrect.

The discovery of several isotopes has only been reported in conference proceedings which are not accepted according to the second criterium. One example from fragmentation experiments why publications in conference proceedings should not be considered is $^{118}$Tc and $^{120}$Ru which had been reported as being discovered in a conference proceeding \cite{1996Cza01} but not in the subsequent refereed publication \cite{1997Ber01}.

The initial literature search was performed using the databases ENSDF \cite{2008ENS01} and NSR \cite{2008NSR01} of the National Nuclear Data Center at Brookhaven National Laboratory. These databases are complete and reliable back to the early 1960's. For earlier references, several editions of the Table of Isotopes were used \cite{1940Liv01,1944Sea01,1948Sea01,1953Hol02,1958Str01,1967Led01}. A good reference for the discovery of the stable isotopes was the second edition of Aston's book ``Mass Spectra and Isotopes'' \cite{1942Ast01}. For the isotopes of the radioactive decay chains several books and articles were consulted, for example, the 1908 edition of ``Gmelin-Kraut's Handbuch der anorganischen Chemie'' \cite{1908Fri01}, Soddy's 1911 book ``The chemistry of the radio-elements'' \cite{1911Sod01}, the 1913 edition of Rutherford's book ``Radioactive substances and their radiations'' \cite{1913Rut01}, and the 1933 article by Mary Elvira Weeks ``The discovery of the elements. XIX. The radioactive elements'' published in the Journal of Chemical Education \cite{1933Wee01}. In addition, the wikipedia page on the radioactive decay chains was a good starting point \cite{2011wik02}.

The isotopes within the radioactive decay chains were treated differently. Their decay properties were largely measured before the concept of isotopes was established. Thus we gave credit to the first observation and identification of a specific activity, even when it was only later placed properly within in the decay chain.

\begin{figure}
	\centering
	\includegraphics[scale=.75]{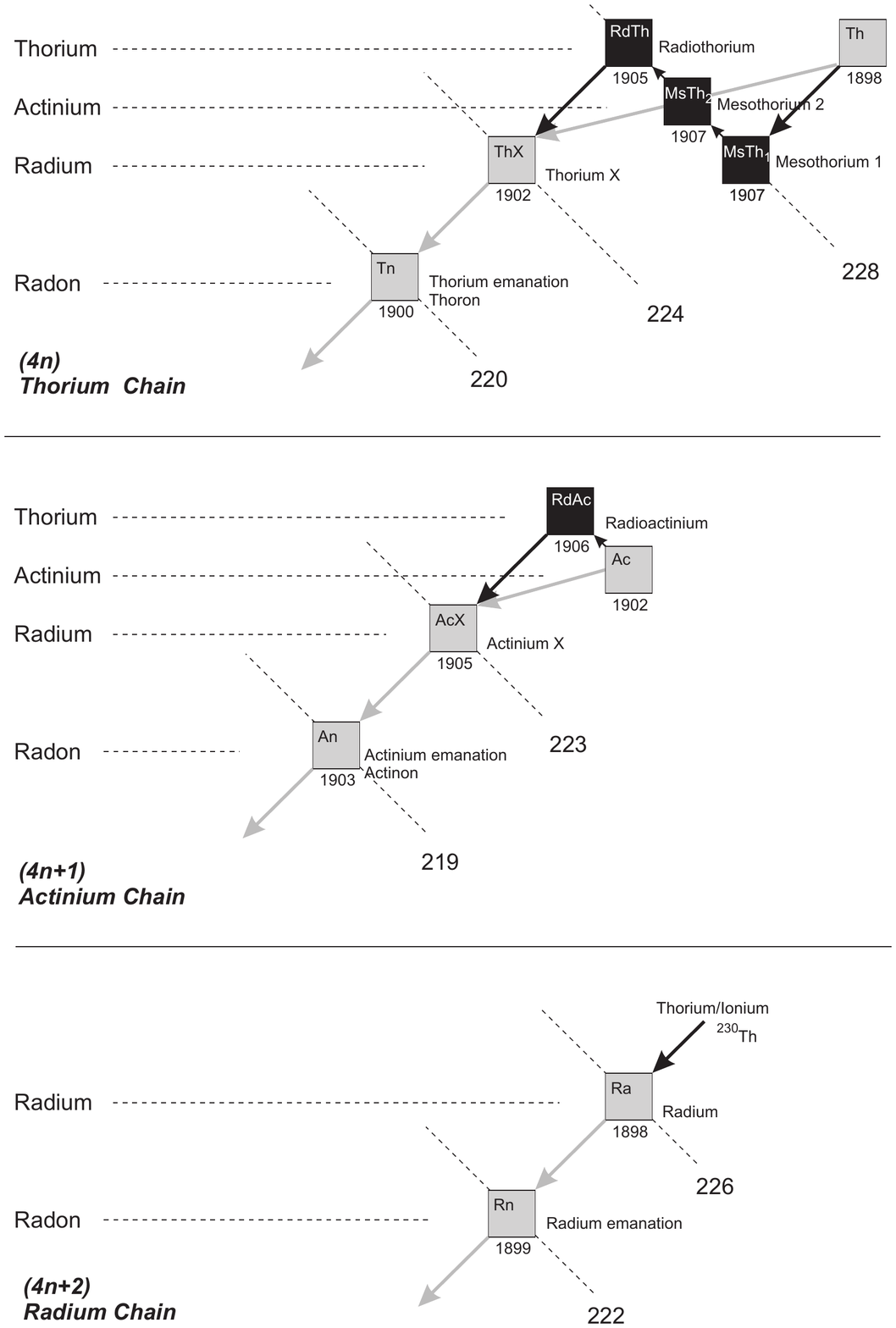}
	\caption{Original nomenclature of radon, radium, actinium, and thorium isotopes within the three natural occurring radioactive decay series. The grey squares connected by the grey arrows depict the activities labeled by Rutherford in his Bakerian lecture \cite{1905Rut01}. The black squares correspond to radioactive substances discovered later.}
\label{f:chain}
\end{figure}

Figure \ref{f:chain} summarizes the isotopes of the three natural occurring radioactive decay series with their original nomenclature. This notation of the original substances introduced by Rutherford during his Bakerian lecture presented on May 19$^{th}$ 1904 and published a year later \cite{1905Rut01} are shown by grey squares and connected by the grey arrows representing $\alpha$ and $\beta$ decay. The decay from actinium to actinium X and from thorium to thorium X was later shown to be more complex. These isotopes are shown as black squares with the corresponding decays shown by black arrows.

%*********************************************************
%*********************************************************

\section{$^{191-229}$At}\vspace{0.0cm}
Astatine was discovered in 1940 by Corson et al.\ by bombarding a bismuth target with $\alpha$ particles \cite{1940Cor03}. A month later Minder reported the observation of element 85 naming it helvetium \cite{1940Min02} which was later shown to be incorrect \cite{1942Kar01}. Also a later claim by Leigh-Smith and Minder naming element 85 anglohelvetium \cite{1942Lei01} was not confirmed. An even earlier report of the discovery of element 85 by Allison et al.\ in 1931 \cite{1931All01} selecting the name alabamine was incorrect \cite{1935Mac01}. The name astatine was officially accepted at the 15$^{th}$ IUPAC conference in Amsterdam in 1949 \cite{1949IUC01}.

\begin{figure}
	\centering
	\includegraphics[scale=.7]{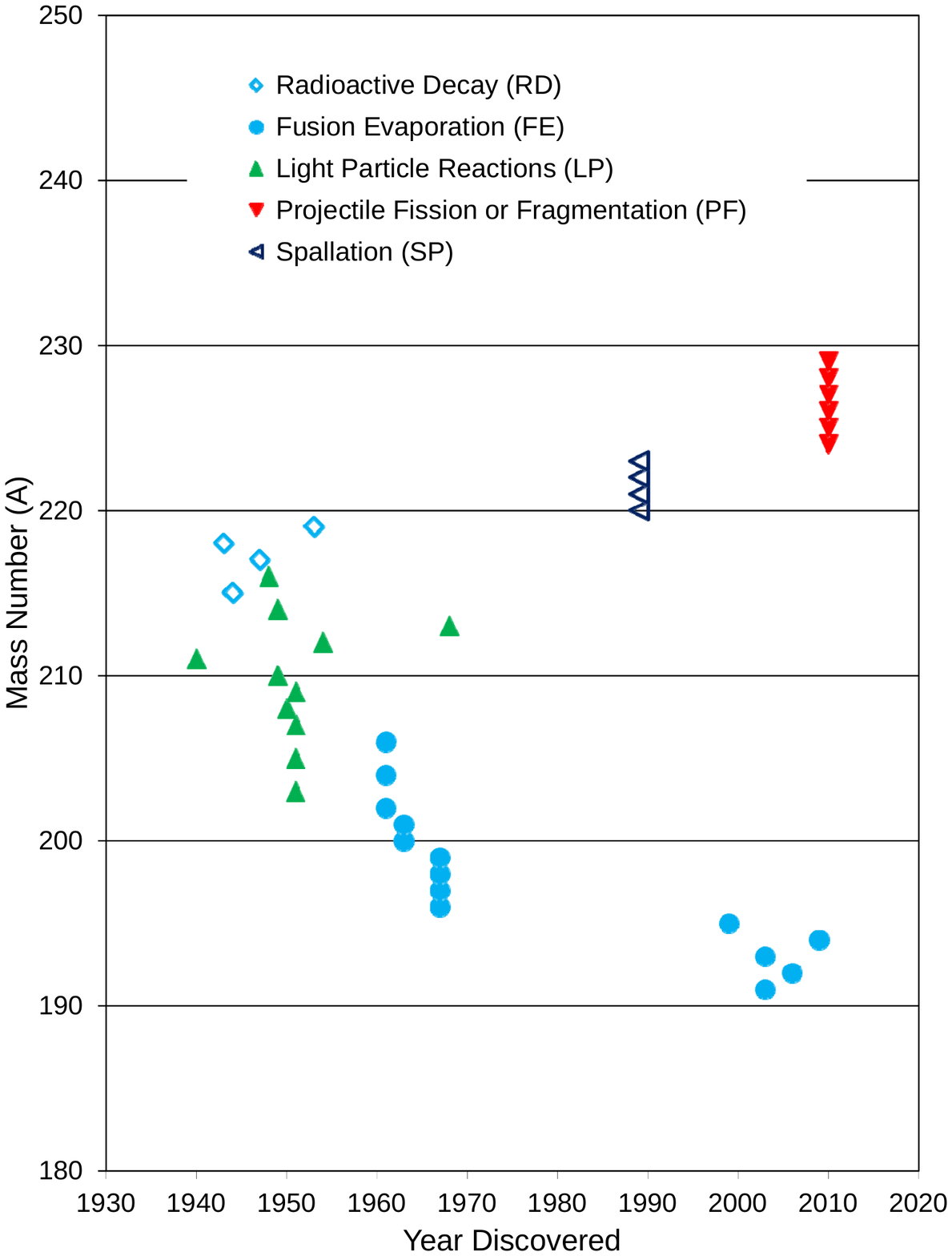}
	\caption{Astatine isotopes as a function of time when they were discovered. The different production methods are indicated.}
\label{f:year-astatine}
\end{figure}

Thirty-nine astatine isotopes from A = 191--229 have been discovered so far and according to the HFB-14 model \cite{2007Gor01} about 37 additional astatine isotopes could exist. Figure \ref{f:year-astatine} summarizes the year of first discovery for all astatine isotopes identified by the method of discovery: radioactive decay (RD), fusion evaporation reactions (FE), light-particle reactions (LP), projectile fission or fragmentation (PF), and spallation (SP). In the following, the discovery of each astatine isotope is discussed in detail and a summary is presented in Table 1. The observation of $^{230}$At was reported in a preprint \cite{2010Ben01}, however, the paper was never accepted for publication in a refereed journal.

\subsection*{$^{191}$At}
Kettunen et al.\ reported the discovery of $^{191}$At in the 2003 paper ``Alpha-decay studies of the new isotopes $^{191}$At and $^{193}$At'' \cite{2003Ket01}. A $^{141}$Pr target was bombarded with 248$-$266~MeV $^{54}$Fe beams from the Jyv\"askyl\"a K-130 cyclotron forming $^{191}$At in (4n) fusion-evaporation reactions. Recoil products were separated with the gas filled recoil separator RITU and implanted into a position sensitive Si detector which also measured subsequent $\alpha$ decay. ``The corresponding mother activity with an alpha-decay energy E$_\alpha$ = 7552(11)~keV and half-life T$_{1/2}$ = (1.7$^{+1.1}_{-0.5}$) ms was assigned to originate from the equivalent 1/2$^+$ state in $^{191}$At...'' The quoted half-life is the currently accepted value.

\subsection*{$^{192}$At}
In the 2006 paper ``$\alpha$-decay spectroscopy of the new isotope $^{192}$At'', Andreyev et al.\ announced the discovery of $^{192}$At \cite{2006And01}. A $^{144}$Sm target was bombarded with a 230~MeV $^{51}$V beam from the GSI UNILAC heavy ion accelerator producing $^{192}$At in the (3n) fusion-evaporation reaction. Recoil products were separated with the velocity filter SHIP and implanted in a 16-strip position-sensitive silicon detector which also measured subsequent $\alpha$ decay. ``Two $\alpha$-decaying isomeric states with half-lives of 88(6)~ms and 11.5(6)~ms were identified in the new isotope $^{192}$At, both of them having complex decay paths to the excited states in the daughter nucleus $^{188}$Bi.'' The quoted half-lives correspond to the currently accepted values for the ground state and an isomeric state, respectively.

\subsection*{$^{193}$At}
Kettunen et al.\ reported the discovery of $^{193}$At in the 2003 paper ``Alpha-decay studies of the new isotopes $^{191}$At and $^{193}$At'' \cite{2003Ket01}. A $^{141}$Pr target was bombarded with 264$-$272~MeV $^{56}$Fe beams from the Jyv\"askyl\"a K-130 cyclotron forming $^{193}$At in (4n) fusion-evaporation reactions. Recoil products were separated with the gas filled recoil separator RITU and implanted into a position sensitive Si detector which also measured subsequent $\alpha$ decay. ``The corresponding mother activity with the alpha-decay energy E$_\alpha$=7295(5)~keV and half-life T$_{1/2}$=(28$^{+5}_{-4}$)~ms was assigned to originate from the equivalent 1/2$^+$ state in $^{193}$At...'' The quoted half-life is the currently accepted value. Previously, the observation of $^{193}$At was reported in a conference proceeding \cite{1995Lei01}.

\subsection*{$^{194}$At}
In 2009 Andreyev et al.\ reported the observation of $^{194}$At in the paper ``$\alpha$ decay of $^{194}$At'' \cite{2009And01}. $^{141}$Pr targets were bombarded with a 259~MeV $^{56}$Fe beam from the GSI UNILAC producing $^{194}$At in (3n) fusion-evaporation reactions. Residues were separated with the velocity filter SHIP and implanted in a 16-strip position-sensitive silicon detector which also recorded subsequent $\alpha$ decay. ``Thus, two different half-life values for decays attributed to $^{194}$At identify two $\alpha$-decaying isomeric states in this nucleus. The 310(8)~ms isomer decaying to $^{190}$Bi$^{m1}$ will further be denoted as $^{194}$At$^{m1}$ while the 253(10)~ms isomer decaying to $^{190}$Bi$^{m2}$ will be denoted as $^{194}$At$^{m2}$.'' These half-lives correspond to the currently accepted values for isomeric states. Previously, a half-life of 180(80)~ms was reported in a conference proceeding \cite{1995Lei01}.

\subsection*{$^{195}$At}
Tagaya et al.\ reported the discovery of $^{195}$At in the 1999 paper ``The $\alpha$-decay energies and halflives of $^{195g,m}$At and $^{199}$Fr'' \cite{1999Tag01}. $^{169}$Tm targets were bombarded with a 215~MeV $^{36}$Ar beam from the RIKEN ring cyclotron to form $^{195}$At in ($\alpha$6n) fusion-evaporation reactions. Recoils were separated with the gas-filled recoil separator GARIS and implanted in a position sensitive detector which also recorded subsequent $\alpha$ decay. ``We therefore assigned the corresponding $\alpha$1 events to the decay of $^{195g}$At, of which the E$_{\alpha}$ and T$_{1/2}$ values were determined to be 7105$\pm$30~keV and 146$^{+21}_{-17}$~ms.'' Tagaya et al.\ also reported an 385$^{+69}_{-51}$~ms isomeric state which is currently assigned the ground state with a half-life of 328(20)~ms. Previously, a half-life of 630$^{+320}_{-160}$~ms was reported in a conference proceeding \cite{1995Lei01}.

\subsection*{$^{196-199}$At}
Treytl and Valli identified $^{196}$At, $^{197}$At, $^{198}$At, and $^{199}$At in the 1967 article ``Alpha decay of neutron deficient astatine isotopes'' \cite{1967Tre01}. Enriched $^{185}$Re and $^{187}$Re targets were bombarded with 100--200~MeV $^{20}$Ne beams from the Berkeley HILAC. Reaction products were collected on a silver foil by a helium jet and rotated in front of a Si(Au) surface barrier detector. ``ASTATINE-199: The peak at 6.638~MeV with a half-life of 7.2~sec clearly belongs to $^{199}$At... ASTATINE-198: The alpha peaks at 6.747~MeV with a half-life of 4.9~sec and at 6.847~MeV with 1.5~sec have excitation functions similar to $^{198}$Po. We have assigned the former one to the ground state and the second one to an isomeric state of $^{198}$At... ASTATINE-197: The peak at 6.957~MeV with a half-life of 0.4~sec belongs to $^{197}$At... ASTATINE-196: In subsequent runs, an alpha peak at 7.055~MeV with a half-life of 0.3~sec was observed. The excitation function shown in [the figure] clearly follows that of $^{196}$Po.'' The measured half-lives of 0.3(1)~s for $^{196}$At, 0.4(1)~s for $^{197}$At, 4.9(5)~s for $^{198}$At, and 7.2(5)~s for $^{199}$At, are close to the currently accepted values of 0.388(7)~s, 0.350(40)~s, 3.8(4)~s, and 7.03(15)~s, respectively.

\subsection*{$^{200-201}$At}
In the 1963 paper ``Alpha decay of neutron-deficient astatine isotopes'', Hoff et al.\ reported the first observation of $^{200}$At and $^{201}$At \cite{1963Hof01}. A gold foil was bombarded with a $^{12}$C beam with energies up to 125~MeV from the Berkeley Hilac. Alpha-particle spectra were measured with a 180$^\circ$ double-focusing spectrograph. ``An $\alpha$-emitter with a half-life of 1.5$\pm$0.1~min and an $\alpha$-particle energy of 6.342$\pm$0.006~MeV has been assigned to $^{201}$At... Two $\alpha$-groups with a half-life 0.9$\pm$0.2~min and energies of 6.412$\pm$0.009 and 6.465$\pm$0.011~MeV have been tentatively assigned to $^{200}$At.'' These half-lives of 0.9(2)~min for $^{200}$At and 1.5(1)~min for $^{201}$At agree with the currently accepted values of 43(1)~s and 85.2(16)~s, respectively. Earlier, Barton et al.\ reported half-lives of 43~s and 1.7~min, but were only able to assign them to astatine isotopes with A$<$202 and A$<$203, respectively \cite{1951Bar01}.

\subsection*{$^{202}$At}
The paper ``$\alpha$-particle branching ratios for neutron-deficient astatine isotopes'' by Latimer et al.\ reported the observation of $^{202}$At in 1961 \cite{1961Lat01}. Gold and platinum foils were irradiated with 50$-$125~MeV $^{12}$C and 65$-$130~MeV $^{14}$N beams, respectively, from the Berkeley HILAC. Alpha-particle spectra were measured with a gridded ionization chamber following chemical separation. ``Using the reported $\alpha$-branching ratio of 0.02 for $^{202}$Po, we have calculated an alpha-branching ratio of 0.120$\pm$0.008 for $^{202}$At, corresponding to a partial $\alpha$-half-life of 25~min... The over-all half-lives observed are in agreement with those reported by Hoff et al.\ \cite{1959Hof01}...'' The overall half-life quoted for $^{202}$At was 3.0(2)~min which agrees with the currently accepted half-life of 184(1)~s. The reference to Hoff et al.\ corresponds to a conference abstract. Hoff et al.\ published their results in a refereed journal two years later \cite{1963Hof01}. Also, about three months later Forsling et al.\ independently reported a 3(1)~min half-life for $^{202}$At \cite{1961For01}.

\subsection*{$^{203}$At}
$^{203}$At was identified by Barton et al.\ and published in the 1951 paper ``Radioactivity of astatine isotopes'' \cite{1951Bar01}. $^{209}$Bi was irradiated with $^4$He beams of up to 380~MeV from the Berkeley 184-in.\ cyclotron. Alpha spectra were recorded with an alpha-pulse analyzer following chemical separation. ``For the present we shall assume the 7-min 6.10-Mev group to be At$^{203}$ and designate the 6.35-Mev group with 1.7-min half-life as At$^{<203}$.'' This value agrees with the currently adopted value of 7.4(2)~min. About three months earlier Miller et al.\ \cite{1950Mil01} measured an 11~min half-life by bombarding a gold target with a $^{13}$C beam and suggested the possibility that they had formed the 7~min $^{203}$At activity based on a private communication with Barton et al.

\subsection*{$^{204}$At}
The paper ``$\alpha$-particle branching ratios for neutron-deficient astatine isotopes'' by Latimer et al.\ reported the observation of $^{204}$At in 1961 \cite{1961Lat01}. Gold and platinum foils were irradiated with 50$-$125~MeV $^{12}$C and 65$-$130~MeV $^{14}$N beams, respectively, from the Berkeley HILAC. Alpha-particle spectra were measured with a gridded ionization chamber following chemical separation. ``In this study, an $\alpha$-group of 5.95~MeV energy and half-life of 9$\pm$1~min has been observed. Excitation functions support the assignment of this activity to $^{204}$At.'' This value agrees with the currently accepted value of 9.12(11)~min. An earlier report of a 22~min half-life \cite{1951Bar01} was evidently incorrect. Also, about three months later Forsling et al.\ independently reported a 9(3)~min half-life \cite{1961For01} and in 1959 Hoff et al.\ had reported a half-life of 9.3(2)~min in a conference abstract \cite{1959Hof01}.

\subsection*{$^{205}$At}
$^{205}$At was identified by Barton et al.\ and published in the 1951 paper ``Radioactivity of astatine isotopes'' \cite{1951Bar01}. $^{209}$Bi was irradiated with $^4$He beams of up to 380~MeV from the Berkeley 184-in.\ cyclotron. Alpha spectra were recorded with an alpha-pulse analyzer following chemical separation. ``For the present, we shall attribute the alpha-particle, which was found to decay with a 25-min half-life to At$^{205}$.'' This value is consistent with the currently adopted value of 26.9(8)~min. About three months earlier Miller et al.\ \cite{1950Mil01} measured an 25~min half-life by bombarding a gold target with a $^{13}$C beam and suggested the possibility that they had formed the 24~min $^{205}$At activity based on a private communication with Barton et al.

\subsection*{$^{206}$At}
The paper ``$\alpha$-particle branching ratios for neutron-deficient astatine isotopes'' by Latimer et al.\ reported the observation of $^{206}$At in 1961 \cite{1961Lat01}. Gold and platinum foils were irradiated with 50$-$125~MeV $^{12}$C and 65$-$130~MeV $^{14}$N beams, respectively, from the Berkeley HILAC. Alpha-particle spectra were measured with a gridded ionization chamber following chemical separation. ``A least-squares analysis of several of the curves for which the statistics were good yielded a value of 29.5$\pm$0.6~min for the half-life of $^{206}$At.'' This value agrees with the currently accepted value of 30.6(8)~min. An earlier report of a 2.6~h half-life \cite{1951Bar01} was evidently incorrect. Also, about three months later Forsling et al.\ independently reported a 20(10)~min half-life \cite{1961For01} and in 1959 Hoff et al.\ had reported a half-life of 31.0(15)~min in a conference abstract \cite{1959Hof01}.

\subsection*{$^{207}$At}
$^{207}$At was identified by Barton et al.\ and published in the 1951 paper ``Radioactivity of astatine isotopes'' \cite{1951Bar01}. $^{209}$Bi was irradiated with $^4$He beams of up to 380~MeV from the Berkeley 184-in.\ cyclotron. Alpha spectra were recorded with an alpha-pulse analyzer following chemical separation. ``At 75~Mev a new activity appeared having a half-life of about 2~hr, and this has been assigned to At$^{207}$ formed by the ($\alpha$,6n) reaction.'' This value agrees with the currently adopted value of 1.81(3)~h.

\subsection*{$^{208}$At}
In 1950 Hyde et al.\ reported the first observation of $^{208}$At in the paper ``Low mass francium and emanation isotopes of high alpha-stability'' \cite{1950Hyd01}. Thorium foils were bombarded with up to 350~MeV protons from the Berkeley 184-inch cyclotron. $^{212}$Fr was chemically separated and $^{208}$At was populated by $\alpha$-decay. Alpha spectra were measured with an ionization chamber. ``High volatility is characteristic of astatine, and this 5.65~Mev activity was judged to be the At$^{208}$ daughter of Fr$^{212}$.'' The measured half-life of 1.7~h agrees with the currently adopted value of 1.63(3)~h.

\subsection*{$^{209}$At}
$^{209}$At was identified by Barton et al.\ and published in the 1951 paper ``Radioactivity of astatine isotopes'' \cite{1951Bar01}. $^{209}$Bi was irradiated with $^4$He beams of up to 380~MeV from the Berkeley 184-in.\ cyclotron. Alpha spectra were recorded with an alpha-pulse analyzer following chemical separation. ``An activity assigned to At$^{209}$ is characterized by a half-life of 5.5$\pm$0.3~hr and an alpha-particle of 5.65~Mev.'' This half-life agrees with the currently adopted value of 5.41(5)~h.

\subsection*{$^{210}$At}
Kelly and Segre first observed $^{210}$At and reported their results in the 1949 paper ``Some excitation functions of bismuth'' \cite{1949Kel01}. Bismuth targets were bombarded with 29~MeV $^4$He beams from the Berkeley 60-inch cyclotron. Resulting activities were measured with a parallel plate ionization chamber. ``Careful investigation, which will be discussed in detail later, showed that the Po$^{210}$ came from the Bi($\alpha$,3n) reaction producing At$^{210}$ which in turn decays to Po$^{210}$ by orbital electron capture, with a half-life of 8.3~hr.'' This value is included in the calculation of the currently accepted half-life of 8.1(4)~h.

\subsection*{$^{211}$At}
The discovery of $^{211}$At was reported in ``Artificially radioactive element 85'' by Corson et al.\ in 1940 \cite{1940Cor01}. The Berkeley 60-inch cyclotron was used to bombard a bismuth target with 32~MeV alpha particles. Alpha particles, gamma-, and x-rays were measured following chemical separation. ``All these radiations separate together chemically as element 85, and all show the same half-life of 7.5~hours. The probable explanation of these effects is the following: Bi$^{209}$, by an ($\alpha$,2n) reaction, goes to 85$^{211}$, which decays either by K-electron capture to actinium C'(Po$^{211}$) or by alpha-particle emission (range 4.5~cm) to Bi$^{207}$.'' The measured half-life agrees with the currently accepted value of 7.214(7)~h. The discovery of the element astatine in this experiment had been published earlier without a mass assignment \cite{1940Cor03}.

\subsection*{$^{212}$At}
Winn reported the observation of $^{212}$At in the 1954 paper ``Short-lived alpha emitters produced by $^3$He and heavy ion bombardments'' \cite{1954Win01}. 28~MeV $\alpha$ particles from the Birmingham cyclotron bombarded a bismuth target forming $^{212}$Bi in the reaction $^{209}$Bi($\alpha$,n). The alpha activity was measured with a zinc sulphide screen attached to a light guide and a magnetically shielded phototube. Results were summarized in a table, quoting an observed half-life of 0.22(3)~s, which is close to the currently accepted value of 0.314(2)~s. Winn did not consider this observation a new discovery referring to the 1948 Table of Isotopes which listed a half-life of 0.25~s based on a private communication \cite{1948Sea01}.

\subsection*{$^{213}$At}
In the 1968 article ``New neptunium isotopes, $^{230}$Np and $^{229}$Np'' Hahn et al.\ reported the observation of $^{213}$At \cite{1968Hah01}. Enriched $^{233}$U targets were bombarded with 32$-$41.6~MeV protons from the Oak Ridge Isochronous Cyclotron forming $^{229}$Np in (p,5n) reactions, respectively. Reaction products were implanted on a catcher foil which was periodically rotated in front of a surface barrier Si(Au) detector. Isotopes populated by subsequent $\alpha$ emission were measured. ``The $\alpha$-particle energies found for the $^{225}$Pa series are more precise than the previously available values: $^{225}$Pa, 7.25$\pm$0.02~MeV (new value); $^{221}$Ac, 7.63$\pm$0.02~MeV; $^{217}$Fr, 8.31$\pm$0.02~MeV and $^{213}$At, 9.06$\pm$0.02~MeV.'' The observation of $^{213}$At was not considered new, referring to an unpublished thesis \cite{1951Key01}. The currently accepted half-life is 125(6)~ms.

\subsection*{$^{214}$At}
Meinke et al.\ reported the observation of $^{214}$At in the 1949 paper ``Three additional collateral alpha-decay chains'' \cite{1949Mei01}. Thorium was bombarded by 150~MeV deuterons from the Berkeley 184-inch cyclotron. Alpha-decay chain from $^{226}$Pa was measured following chemical separation. ``Although the mass type has not yet been identified through known daughters as above, general considerations with regard to the method of formation and half-life of the parent substance, and the energies of all the members of the series suggest a collateral branch of the 4n+2 family: $_{91}$Pa$^{226}\stackrel{\alpha}{\longrightarrow}_{89}$Ac$^{222}\stackrel{\alpha}{\longrightarrow}_{87}$Fr$^{218}\stackrel{\alpha}{\longrightarrow}_{85}$Fr$^{214}\stackrel{\alpha}{\longrightarrow}_{83}$Bi$^{210}$(RaE).'' In a table summarizing the energies and half-lives of the decay chain only the $\alpha$-decay energy was given for $^{214}$At stating a calculated half-life of 10$^{-6}$~s. The currently accepted half-life is 558(10)~ns.

\subsection*{$^{215}$At}
In the 1944 paper ``Das Element 85 in der Actiniumreihe'', Karlik and Bernert reported the first observation of $^{215}$At \cite{1944Kar01}. The range of $\alpha$ particles from a actinium emanation source was measured with an ionization chamber. ``Wir fanden in einem Verh\"altnis von ungef\"ahr 5$\cdot$10$^{-6}$ zur Actinium A-Strahlung eine $\alpha$-Strahlung mit einer extrapolierten Reichweite von 8,0~cm (15$^\circ$, 760~mm), was 8,4 MeV Zerfallsenergie entspricht. Dieser Betrag steht in sehr guter \"Ubereinstimmung mit dem Wert, der sich ergibt, wenn man in dem Diagramm der Zerfallsenergie als Funktion der Massenzahl die Kurve von der Ordnungszahl 85 bis zur Massenzahl 215 extrapoliert, so da\ss\ uns die Zuordnung der neuen $\alpha$-Strahlung zu dem Kern 85$^{215}$ (entstanden aus Ac A durch $\beta$-Zerfall) berechtigt erscheint.'' [We found an $\alpha$ radiation with a ratio of approximately 5$\cdot$10$^{-6}$ relative to the actinium A radiation which has an extrapolated range of 8.0~cm (15$^\circ$, 760~mm), corresponding to a decay energy of 8.4~MeV. This value agrees very well with the extrapolated value for mass number 215 in a plot of the decay energy as a function of the mass number for atomic number 85. Thus it is reasonable to assign the new $\alpha$ radiation to the nuclide 85$^{215}$ (produced by $\beta$ decay from Ac A).] The presently adopted half-life is 100(20)~$\mu$s.

\subsection*{$^{216}$At}
In ``Artificial collateral chains to the thorium and actinium families,'' Ghiorso et al.\ discovered $^{216}$At in 1948 \cite{1948Ghi01}. Thorium targets were irradiated with 80~MeV deuterons from the Berkeley 184-inch cyclotron. The $\alpha$-decay chain beginning at $^{228}$Pa was measured following chemical separation. ``After the decay of the above-described series, a second group of alpha-particle emitters can be resolved. This second series, which decays with the 22-hour half-life of its protactinium parent, is a collateral branch of the 4n radioactive family as follows: $_{91}$Pa$^{228}\stackrel{\alpha}{\longrightarrow}_{89}$Ac$^{224}\stackrel{\alpha}{\longrightarrow}_{87}$Fr$^{220}\stackrel{\alpha}{\longrightarrow}_{85}$At$^{216}\stackrel{\alpha}{\longrightarrow}$...''
The measured half-life of about 10$^{-3}$~s is consistent with the presently adopted value of 0.3~ms. In 1940, Minder \cite{1940Min02} and later in 1942, Leigh-Smith and Minder \cite{1942Lei01} had reported the observation of $^{216}$At $\beta$-decay which was evidently incorrect \cite{1942Kar01}. Also the observation of $^{216}$At in the natural thorium radioactive decay chain \cite{1943Kar01} was not correct \cite{1950Per01}.

%In``Ein weiterer dualer Zerfall in der Thoriumreihe,'' Karlik and Bernert discovered $^{216}$At in 1943 \cite{1943Kar02}. The range of $\alpha$ particles from a thorium emanation source was measured with an ionization chamber. ``Unter der Voraussetzung, da\ss\ es sich hier, analog dem Radium A, um einen $\beta$-Zerfall von Thorium A handelt, der zu dem $\alpha$-strahlenden Isotop 216 des Elementes 85 f\"uhrt, errechnet sich ein Abzweigungsverh\"altnis von 1,35$\cdot$l0$^{-4}$ gegen\"uber dem $\alpha$-Zerfall; dabei wurde auf die Thorium A-Menge aus der Zahl der nach l\"angerer Zeit vorhandenen Thorium C'-Strahlen geschlossen... Aus der Reichweite der $\alpha$-Strahlen berechnet man eine Zerfallsenergie von 7,76 MeV.'' [If - in analogy to radium A - the decay corresponds to $\beta$ decay from thorium A leading to an $\alpha$ emitting isotope of element 85, the branching ratio could be calculated to be 1.35$\cdot$10$^{-4}$ relative to $\alpha$ emission; the amount of thorium A was calculated from the thorium C' radiation which accumulated over a longer time... From the range of the $\alpha$ particles a decay energy of 7.76~MeV is calculated.] A year earlier Leigh-Smith and Minder had reported the observation of $^{216}$At $\beta$-decay \cite{1942Lei01} which was evidently incorrect.  Leigh-Smith and Minder, apparently unaware of the work by Corson et al. \cite{1940Cor03} who had discovered astatine two years earlier, suggested to name the new element anglo-helvetium \cite{1942Lei01}.

\subsection*{$^{217}$At}
Hagemann et al.\ discovered $^{217}$At in 1947 in ``The (4n+1) radioactive series: the decay products of U$^{233}$'' \cite{1947Hag01}. The half-lives and $\alpha$- and $\beta$-decay energies of the nuclides in the decay chain of $^{233}$U were measured. ``These decay products, which constitute a substantial fraction of the entire missing, 4n+1, radioactive series are listed together with their radioactive properties, in [the table].'' The measured half-life of 18~ms is within a factor of two of the presently accepted value of 32.3(4)~ms. Hagemann et al.\ acknowledge the simultaneous observation of $^{217}$At by English et al.\ which was submitted only a day later and published in the same issue of Physical Review on the next page \cite{1947Eng01}.

\subsection*{$^{218}$At}
$^{218}$At was identified by Karlik and Bernert in the 1943 paper ``Eine neue nat\"urliche $\alpha$-Strahlung'' \cite{1943Kar01}. The range of $\alpha$ particles from a radium A source was measured with an ionization chamber. ``Eine $\beta$-Umwandlung von Radium A w\"urde zu einem Isotop des
Elementes 85 von der Massenzahl 218 f\"uhren... Die entsprechende Energie betr\"agt 6.6$_3$ MeV, bzw. die gesamte Zerfallsenergie 6,7$_5$ MeV. Aus der Geiger-Nutallschen Beziehung w\"urde sich daraus eine Halbwertszeit in der Gro\ss enordnung von Sekunden ableiten, was mit unseren Beobachtungen im Einklang steht.'' [A potential $\beta$ decay of radium A would lead to an isotope of element 85 with a mass number of 218... The corresponding energy is 6.6$_3$ MeV, corresponding to a total decay energy of 6.7$_5$ MeV. From this energy a half-life on the order of seconds can be derived from the Geiger-Nutall relation which is consistent with our observations.] The currently adopted half-life for $^{218}$At is 1.5(3)~s.

\subsection*{$^{219}$At}
In 1953 $^{219}$At was first reported by Hyde and Ghiorso in ``The alpha-branching of AcK and the presence of astatine in nature'' \cite{1953Hyd01}. A 20-mC $^{227}$Ac source was used to study the nuclide of the 4n+3 decay series by chemical and physical separation and measuring the radioactivity with an alpha-ray differential pulse analyzer. ``The observed branching rate is ca 4$\times$10$^{-5}$, and the At$^{219}$ daughter decays predominantly by the emission of 6.27 Mev alpha-particles with a half-life of 0.9 minute to the new isotopes Bi$^{215}$, which in turn emits $\beta^-$ particles with a half-life of 8 minutes.'' The measured half-life of 0.9~min for $^{219}$At is included in the calculated average of the currently adopted value of 56(3)~s.

\subsection*{$^{220}$At}
In 1989 Liang et al.\  reported the first observation of $^{220}$At in ``A new isotope $_{85}^{220}$At$_{135}$'' \cite{1989Lia01}. Thorium oxide was bombarded with 200~MeV protons from the Orsay synchrocyclotron. $^{220}$At was separated with the ISOCELE II on-line mass separator and transported to a measuring station consisting of a 4$\pi$ $\beta$-detector and two Ge(Li) detectors. ``A new isotope $^{220}$At has been identified among the mass-separated products of a spallation reaction of ThO$_2$. Its half-life has been found to be 3.71$\pm$0.04~min.'' This half-life is the currently adopted value. Less than three months later, Burke et al.\ independently reported a half-life of 3.73(13)~min \cite{1989Bur01}.

\subsection*{$^{221-223}$At}
In the 1989 paper ``New neutron-rich isotopes of astatine and bismuth'' Burke et al.\ described the observation of $^{221}$At, $^{222}$At and $^{223}$At  \cite{1989Bur01}. A thorium/tantalum metal-foil target was bombarded with 600~MeV protons from the CERN synchro-cyclotron. Astatine isotopes were produced in spallation reactions and separated with the ISOLDE-II on-line separator. Beta-ray spectra were measured with a 4$\pi$ plastic scintillator. ``Multiscaling of the 4$\pi$ plastic scintillator signal gave a half-life of 2.3(2) min. This can be assigned to $^{221}$At... At mass A=222, an activity with a half-life of 54(10)~s has been observed and assigned to $^{222}$At... The most neutron-rich astatine isotope seen in the present experiment was $^{223}$At. Its half-life was measured to be 50(7)~s.'' The measured half-lives of 2.3(2)~min, 54(10)~s, and 50(7)~s for $^{221}$At, $^{222}$At and $^{223}$At, respectively, are the currently accepted values.

\subsection*{$^{224}$At}
In the 2010 paper ``Discovery and investigation of heavy neutron-rich isotopes with time-resolved Schottky spectrometry in the element range from thallium to actinium'', Chen et al.\ described the discovery of $^{224}$At \cite{2010Che01}. A beryllium target was bombarded with a 670~MeV/u $^{238}$U beam from the GSI heavy-ion synchrotron SIS and projectile fragments were separated with the fragment separator FRS. The mass and half-life of $^{224}$At was measured with time-resolved Schottky Mass Spectrometry in the storage-cooler ring ESR. ``In [the figure] time traces and their projection into a frequency spectrum are shown for the new isotope $^{224}$At and close-lying ions.'' The quoted half-life of 76$^{+138}_{-23}$~s is currently the only measured value for $^{224}$At.

\subsection*{$^{225-229}$At}
$^{225}$At, $^{226}$At, $^{227}$At, $^{228}$At, and $^{229}$At were discovered by Alvarez-Pol and the results were published in the 2010 paper ``Production of new neutron-rich isotopes of heavy elements in fragmentation reactions of $^{238}$U projectiles at 1A GeV'' \cite{2010Alv01}. A beryllium  target was bombarded with a 1~A GeV $^{238}$U beam from the GSI SIS synchrotron. The isotopes were separated and identified with the high-resolving-power magnetic spectrometer FRS. ``To search for new heavy neutron-rich nuclei, we tuned the FRS magnets for centering the nuclei $^{227}$At, $^{229}$At, $^{216}$Pb, $^{219}$Pb, and $^{210}$Au along its central trajectory. Combining the signals recorded in these settings of the FRS and using the analysis technique previously explained, we were able to identify 40 new neutron-rich nuclei with atomic numbers between Z=78 and Z=87; $^{205}$Pt, $^{207-210}$Au, $^{211-216}$Hg, $^{214-217}$Tl, $^{215-220}$Pb, $^{219-224}$Bi, $^{223-227}$Po, $^{225-229}$At, $^{230,231}$Rn, and $^{233}$Fr.''

%*********************************************************
%*********************************************************

\section{$^{193-231}$Rn}\vspace{0.0cm}
Although it is generally accepted that the element radon was discovered by Dorn in 1900 the references to the original papers is not straightforward \cite{1957Par01}. Also arguments have been made to credit Rutherford \cite{2003Mar01} or M. and P. Curie \cite{2000Bre01} with the discovery of radon. In 1923, the International Union of Pure and Applied Chemistry (IUPAC) named the three known emanations - radium, actinium, and thorium - radon (Rn), actinon (An), and thoron (Tn), respectively. Ramsay and Gray had suggested the name niton for radium emanation \cite{1910Ram01} in 1910. Later radon (Rn) was adopted for all radon isotopes, however, the name emanation with the symbol Em was commonly in use for a long time, see for example reference \cite{1967Val01}.

\begin{figure}
	\centering
	\includegraphics[scale=.7]{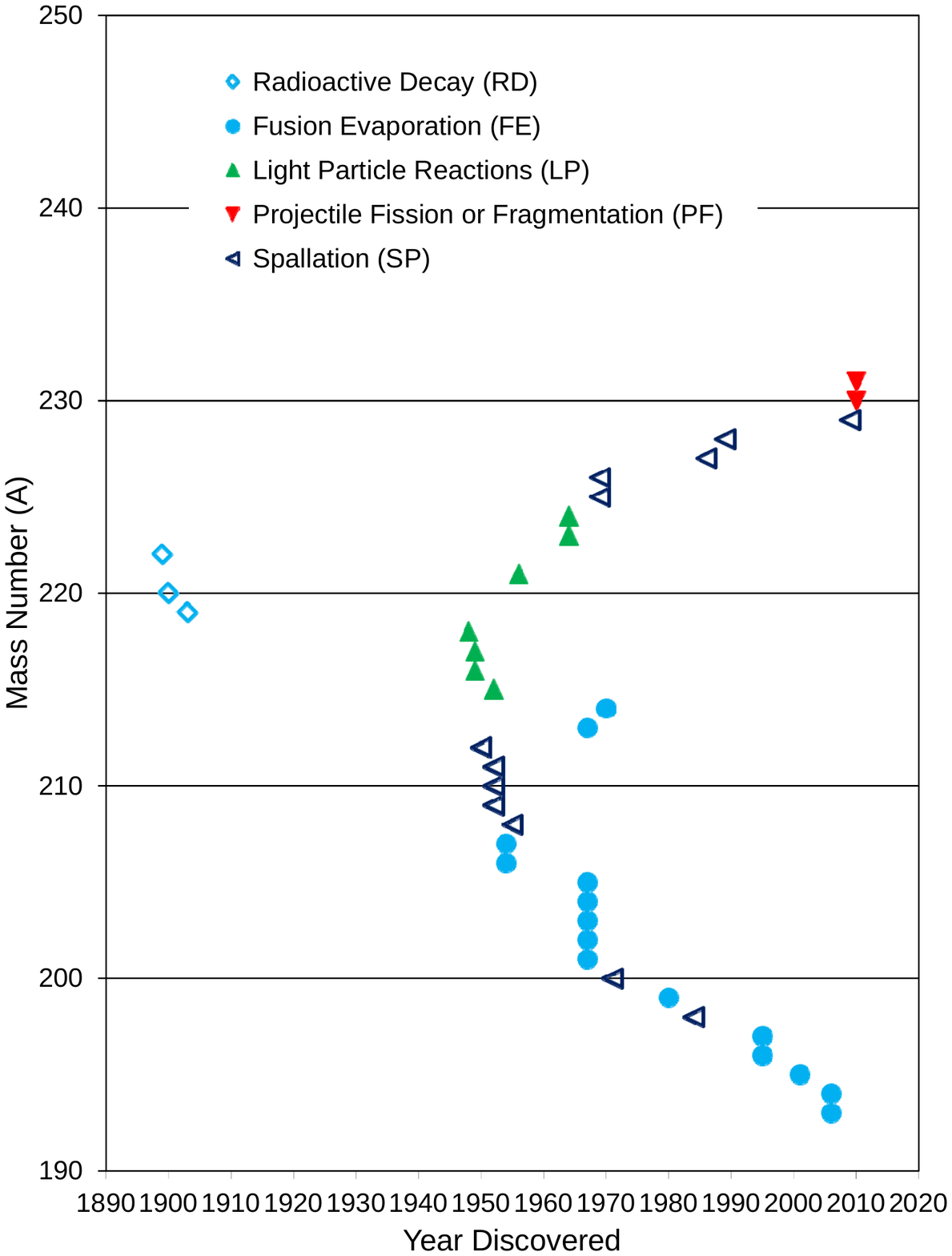}
	\caption{Radon isotopes as a function of time when they were discovered. The different production methods are indicated.}
\label{f:year-radon}
\end{figure}

Thirty-nine radon isotopes from A = 193--231 have been discovered so far and according to the HFB-14 model \cite{2007Gor01} about 48 additional radon isotopes could exist. Figure \ref{f:year-radon} summarizes the year of first discovery for all radon isotopes identified by the method of discovery: radioactive decay (RD), fusion evaporation reactions (FE), light-particle reactions (LP), projectile fission or fragmentation (PF), and spallation (SP). In the following, the discovery of each radon isotope is discussed in detail and a summary is presented in Table 1. The observation of $^{232}$Rn was reported in a preprint \cite{2010Ben01}, however, the paper was never accepted for publication in a refereed journal.

\subsection*{$^{193,194}$Rn}
Andreyev et al.\ reported the first observation of $^{193}$Rn and $^{194}$Rn in the 2006 paper ``$\alpha$ decay of the new isotopes $^{193,194}$Rn'' \cite{2006And02}. A $^{144}$Sm target was bombarded with 231$-$252~MeV $^{52}$Cr beams from the GSI UNILAC forming $^{193}$Rn and $^{194}$Rn in the (3n) and (2n) fusion-evaporation reactions, respectively. Recoil products were separated with the velocity filter SHIP and implanted into a position-sensitive silicon detector which also recorded subsequent $\alpha$ decay. ``By using all 26 full-energy correlated recoil-$\alpha_1$ decays a half-life of T$_{1/2}$=0.78(16)~ms was deduced for $^{194}$Rn... The half-life of T$_{1/2}$($^{193}$Rn)=1.15(27)~ms was deduced from 19 full-energy recoil-$\alpha_1$(7670~keV-7890~keV) decays, which includes 16 events with the full-energy deposition in the PSSD and 3 events in which the energy was shared between the PSSD and BOX detectors.'' Both of these half-lives are the currently adopted values.

\subsection*{$^{195}$Rn}
The discovery of $^{195}$Rn by Kettunen et al.\ was reported in the 2001 paper ``$\alpha$ decay studies of the nuclides $^{195}$Rn and $^{196}$Rn'' \cite{2001Ket01}. A $^{142}$Nd target was bombarded with 239$-$267~MeV $^{56}$Fe beams from the Jyv\"askyl\"a K-130 cyclotron producing $^{195}$Rn  in the (3n) fusion-evaporation reaction. Recoil products were separated with the gas-filled recoil separator RITU and implanted into a position sensitive silicon detector which also measured subsequent $\alpha$ decay. ``Two $\alpha$ decaying isomeric states, with E$_\alpha$=7536(11)~keV [T$_{1/2}$=(6$^{+3}_{-2}$)~ms] for the ground state and E$_\alpha$=7555(11)~keV [T$_{1/2}$=(5$^{+3}_{-2}$)~ms] for an isomeric state were identified in $^{195}$Rn.'' These half-lives are the currently accepted values.

\subsection*{$^{196,197}$Rn}
In the 1995 article ``New $\alpha$-decaying neutron deficient isotopes $^{197}$Rn and $^{200}$Fr,'' Morita et al.\ announced the identification of $^{196}$Rn and $^{197}$Rn \cite{1995Mor01}. A 273.6~MeV $^{36}$Ar beam from the RIKEN ring cyclotron bombarded an enriched $^{166}$Er target forming $^{196}$Rn and $^{197}$Rn in (6n) and (5n) fusion-evaporation reactions, respectively. Reaction products were separated with the gas-filled recoil separator GARIS and implanted in a position-sensitive silicon detector which also measured subsequent $\alpha$ decay. ``The $\alpha$-decay energies (half-lives) of $^{197}$Rn, $^{197m}$Rn and $^{200}$Fr have been determined to be 7261$\pm$30~keV (51$^{+35}_{-15}$~ms), 7370$\pm$30~keV (18$^{+9}_{-5}$~ms), and 7500$\pm$30~keV, (570$^{+270}_{-140}$~ms), respectively.'' Only one $\alpha$-decay event was observed for $^{196}$Rn with 5~ms between the implant and the $\alpha$ particle. The same group reported the half-life of $^{196}$Rn as 3$^{+7}_{-2}$~ms which agrees with the presently accepted value of 4.4$^{+1.3}_{-0.9}$~ms a year later \cite{1997Pu01}. The measured half-life of 51$^{+35}_{-15}$~ms for $^{197}$Rn agrees with the present value of 65$^{+25}_{-14}$~ms. Three months later Enquist et al.\ \cite{1996Enq01} independently reported the observation of the isomeric state which agreed with the value of Morita et al.

\subsection*{$^{198}$Rn}
The discovery of $^{198}$Rn was published in the 1984 paper ``Alpha decay of $^{198}$Rn'' by Calaprice et al.\ \cite{1984Cal01}. Thorium hydroxide targets were bombarded with 600~MeV protons from the CERN synchrocyclotron forming $^{198}$Rn in spallation reactions. Decay curves of $^{198}$Rn were measured following isotope separation with the online mass separator ISOLDE. ``The new nuclide $^{198}$Rn was found to have an $\alpha$-decay energy of 7196$\pm$10~keV and a half-life of 50$\pm$9~ms.'' This half-life agrees with the currently accepted value of 65(3)~ms.

\subsection*{$^{199}$Rn}
In 1980, DiRienzo et al.\ reported the observation of $^{199}$Rn in ``New isotope $^{199}$Rn and evidence for an isomeric state $^{199}$Rn$^m$'' \cite{1980DiR01}. A 200~MeV $^{35}$Cl beam from the BNL three-stage Tandem Accelerator bombarded a $^{169}$Tm target forming $^{199}$Rn in the (5n) fusion-evaporation reaction. Recoil products were separated with a zero-degree recoil separator and implanted in a surface barrier detector which also measured subsequent $\alpha$ decay. ``The other two lines at 6.990$\pm$0.015~MeV and 7.060$\pm$0.012~MeV are assigned to a new isotope $^{199}$Rn.'' The currently accepted half-life is 0.59(3)~s.

\subsection*{$^{200}$Rn}
Hornshoj et al.\ reported the identification of $^{200}$Rn in ``Alpha decay of neutron-deficient radon and polonium isotopes'' in 1971 \cite{1971Hor01}.  Th(OH)$_4$ targets were bombarded with 600~MeV protons from the CERN synchrocyclotron forming $^{200}$Rn in spallation reactions. Alpha-decay spectra were measured following isotope separation with the online mass separator ISOLDE. ``$^{200}$Rn decays by an $\alpha$-group of energy 6.909$\pm$0.008~MeV, see [the figure]. The half-life was found to be 1.0$\pm$0.2~s.'' This value is included in the calculation of the currently accepted value.

\subsection*{$^{201-205}$Rn}
Valli et al.\ reported the discovery of $^{201}$Rn, $^{202}$Rn, $^{203}$Rn, $^{204}$Rn, and $^{205}$Rn in the 1967 article ``Alpha-decay properties of neutron-deficient isotopes of emanation'' \cite{1967Val01}. Platinum, gold, mercury, and thallium targets were bombarded with $^{16}$O, $^{14}$N, and $^{12}$C beams from the Berkeley HILAC. Alpha-particle spectra were measured with a Si(Au) detector following chemical separation. ``Emanation-201: ...The most prominent of the groups, at 6.768~MeV, had a half-life of 3$\pm$1~sec. We tentatively assign it to $^{201}$Em on the following incomplete evidence... Emanation-202: ...By examination of several spectra taken at 15-sec intervals, the half-life was determined to be 13$\pm$2 sec... The excitation function leads to a mass assignment of 202... Emanation-203 and Emanation-203m: ...We assign the 45-sec 6.497~MeV activity to the ground state of $^{203}$Em and the 28-sec 6.547~MeV activity to an isomeric state as this choice fits best in the energy-versus-mass-number curve... Emanation-204: An $\alpha$ activity at 6.416 MeV with a half-life of 75$\pm$2 sec was prominent in the emanation fraction from gold targets bombarded with $^{14}$N or $^{16}$O nuclei or from platinum targets bombarded with $^{16}$O nuclei... the agreement of the $\alpha$ energy with the approximate value to be expected from systematic trends in $\alpha$-decay energies confirm the assignment of the new activity to $^{204}$Em... Emanation-206 and Emanation-205: ...From an analysis of many decay curves of the 6.260-MeV $\alpha$ group we found a two-component mixture with half-life periods of 6.5$\pm$1~min and 1.8$\pm$0.5~min. The longer-lived component corresponds to the $^{206}$Em reported by Stoner and Hyde \cite{1957Sto01}. The 1.8-min period can be assigned to the previously unknown $^{205}$Em from arguments based on our excitation function results.'' The measured half-lives of 3.0(15)~s ($^{201}$Rn), 13(2)~s ($^{202}$Rn), 45(5)~s ($^{203}$Rn), 75(2)~s ($^{204}$Rn), and 1.8(5)~min ($^{205}$Rn) agree with the presently adopted values of 3.8(1)~s, 9.7(1)~s, 44(2)~s, 74.5(14)~s, and 170(4)~s, respectively. The value for $^{201}$Rn corresponds to an isomeric state. Earlier, Stoner and Hyde had reported a 3~min half-life and assigned it to either $^{204}$Rn or $^{205}$Rn \cite{1957Sto01}.

\subsection*{$^{206,207}$Rn}
In the 1954 paper ``The $\alpha$-activity induced in gold by bombardment with nitrogen ions,'' Burcham described the identification of $^{206}$Rn and $^{207}$Rn \cite{1954Bur01}. Gold foils were bombarded with a 75--120~MeV nitrogen beam from the Birmingham Nuffield 60-inch cyclotron forming $^{206}$Rn and $^{207}$Rn in the fusion-evaporation reactions $^{197}$Au($^{14}$N,5n) and $^{197}$Au($^{14}$N,4n), respectively. Alpha-decay curves of the irradiated samples were measured with an ionization chamber. ``Assignment of the 6.25~MeV group of $\alpha$-particles to $^{206}$Em is based on predictions from $\alpha$-decay systematics... The 6.09~MeV group of $\alpha$-particles could come from $^{207}$Em according to the systematics.'' The measured half-lives of 6.5(5)~min for $^{206}$Rn and 11.0(10)~min for $^{207}$Rn are close to the currently adopted values of 5.67(17)~min and 9.25(17)~min, respectively.

\subsection*{$^{208}$Rn}
Momyer and Hyde reported the observation of $^{208}$Rn in the 1955 article ``The influence of the 126-neutron shell on the alpha-decay properties of the isotopes of emanation, francium, and radium'' \cite{1955Mom01}. Thorium foils were bombarded with 340~MeV protons from the Berkeley 184-inch cyclotron. Alpha-particle spectra and decay curves were measured with an ionization chamber following chemical separation. ``In summary, Em$^{208}$ appears to be a 23$\pm$2-minute activity with alpha-particle energy 6.141~MeV.'' This half-life agrees with the currently adopted value of 24.35(14)~min. In a companion paper actually submitted a day earlier, Momyer et al.\ measured the $\alpha$-decay energies in a magnetic spectrograph \cite{1955Mom02}.

\subsection*{$^{209-211}$Rn}
Momyer et al.\ identified $^{209}$Rn, $^{210}$Rn, and $^{211}$Rn in ``Recent studies of the isotopes of emanation, francium and radium'' in 1952 \cite{1952Mom01}. Thorium targets were bombarded with 340~MeV protons from the Berkeley 184-inch cyclotron. Alpha-decay spectra were measured following chemical separation. Results were summarized in a table, assigning half-lives of 31~min, 2.7~h, and 16~h to $^{209}$Rn, $^{210}$Rn, and $^{211}$Rn, which agree with the currently accepted half-lives of 28.5(10)~min, 2.4(1)~h, and 14.6(2)~h, respectively. Half-lives of 23~min and 2.1~h had been previously reported without firm mass assignments \cite{1949Ghi01}.

\subsection*{$^{212}$Rn}
In 1950 Hyde et al.\ reported the first observation of $^{212}$Rn in the paper ``Low mass francium and emanation isotopes of high alpha-stability'' \cite{1950Hyd01}. Thorium foils were bombarded with up to 350~MeV protons from the Berkeley 184-inch cyclotron. $^{212}$Fr was chemically separated and $^{212}$Rn was populated by electron capture. Alpha spectra were measured with an ionization chamber. ``Em$^{212}$ is shown to be a 23-minute alpha-emitter.'' This agrees with the currently adopted half-life of 23.9(12)~min. The same group had reported this activity previously without a mass assignment \cite{1949Ghi01}.

\subsection*{$^{213}$Rn}
Rotter et al.\ observed $^{213}$Rn in 1967 and reported their results in the paper ``The new isotope Ac$^{216}$'' \cite{1967Rot01}. A 78~MeV $^{12}$C beam from the Dubna 1.5~m cyclotron bombarded a lead target forming radium in (xn) reactions. $^{213}$Rn was populated by $\alpha$ decay of $^{217}$Ra. Recoil nuclei were collected on an aluminum foil and $\alpha$-particle spectra were measured with a silicon surface barrier detector. ``We obtained the following $\alpha$-particle energies: Rn$^{213}$ - 8.14~MeV, Fr$^{214}$ - 8.53~MeV, and Ra$^{215}$ - 8.73~MeV.'' Rotter et al.\ did not consider this observation a new discovery referring to a conference abstract \cite{1962Gri01}.

\subsection*{$^{214}$Rn}
In 1970 Torgerson and MacFarlane reported the first observation of $^{214}$Rn in ``Alpha decay of the $^{221}$Th and $^{222}$Th decay chains'' \cite{1970Tor01}. A 10.6~MeV/nucleon $^{16}$O beam from the Yale heavy ion accelerator was used to bombard a $^{208}$Pb target forming $^{222}$Th in (2n) fusion-evaporation reactions. $^{214}$Rn was then populated by subsequent $\alpha$ decays. Recoil products were transported to a stainless steel surface with a helium jet and $\alpha$ spectra were measured with a Si(Au) surface barrier detector. ``However, at $^{16}$O incident energies below 80~MeV, the 9.040~MeV group could be clearly resolved as shown in [the figure].'' Only three days later Valli et al.\ submitted their measurement of a 9.035(10)~MeV $\alpha$ energy assigned to $^{214}$Rn with a 0.27(2)~$\mu$s half-life \cite{1970Val01}. Earlier, the assignment of a 11.7~MeV $\alpha$ energy to $^{214}$Rn \cite{1962Kar01} was evidently incorrect.

\subsection*{$^{215}$Rn}
In 1952, $^{215}$Rn was discovered by Meinke et al.\ and the results were reported in the paper ``Further work on heavy collateral radioactive chains'' \cite{1952Mei01}. Thorium nitrate targets were irradiated with a $^4$He beam from the Berkeley 184-inch cyclotron. $^{227}$U was chemically separated and the decay and energy of $\alpha$-particles were measured with an alpha-particle pulse analyzer. ``An additional short-lived chain collateral to the actinium (4n+3) natural radioactive family has also been partially identified. This chain decays as follows: U$^{227}\rightarrow$Th$^{223}\rightarrow$Ra$^{219}\rightarrow$Em$^{215}\rightarrow$Po$^{211}\rightarrow$Pb$^{207}$.'' An $\alpha$ energy of 8.6(1)~MeV was assigned to $^{215}$Rn. The presently adopted half-life os 2.3(1)~$\mu$s.

\subsection*{$^{216,217}$Rn}
Meinke et al.\ reported the observation of $^{216}$Rn and $^{217}$Rn in the 1949 paper ``Three additional collateral alpha-decay chains'' \cite{1949Mei01}. Thorium was bombarded with 100$-$120~MeV $^4$He beams from the Berkeley 184-inch cyclotron. Alpha-decay chains from $^{228}$U and $^{229}$U were measured following chemical separation. ``The irradiation of thorium with 100-Mev helium ions resulted in the observation of the following collateral branch of the artificial 4n$+$1, neptunium, radioactive family shown with Po$^{213}$ and its decay products: $_{92}$U$^{229}\overset{\alpha}{\rightarrow}_{90}$Th$^{225}\overset{\alpha}{\rightarrow}_{88}$Ra$^{221}\overset{\alpha}{\rightarrow}_{86}$Em$^{217}\ldots$ Immediately after 120-Mev helium ion bombardment of thorium the uranium fraction contains another series of five alpha-emitters, which is apparently a collateral branch of the 4n family: $_{92}$U$^{228}\overset{\alpha}{\rightarrow}_{90}$Th$^{224}\overset{\alpha}{\rightarrow}_{88}$Ra$^{220}\overset{\alpha}{\rightarrow}_{86}$Em$^{216}\ldots$'' In a table summarizing the energies and half-lives of the decay chain only the $\alpha$-decay energies were given for $^{216}$Rn and $^{217}$Rn stating calculated half-lives of $\sim$10$^{-5}$~s and $\sim$10$^{-3}$~s, respectively. The currently accepted half-lives of $^{216}$Rn and $^{217}$Rn are 45(5)~$\mu$sand 540(50)~$\mu$s, respectively.

\subsection*{$^{218}$Rn}
Studier and Hyde announced the discovery of $^{218}$Rn in the 1948 paper ``A new radioactive series - the protactinium series'' \cite{1948Stu01}. Thorium metal targets were bombarded with 19~MeV deuterons and a 38~MeV $^4$He beam from the Berkeley 60-inch cyclotron forming $^{230}$Pa in (d,4n) and ($\alpha$,p5n) reactions. $^{218}$Rn was populated by subsequent $\alpha$ decay after the initial $\beta^-$ decay of $^{230}$Pa to $^{230}$U. Alpha-decay spectra were measured following chemical separation. ``[The figure] shows the frequency distribution of the observed time intervals after correction for random events. The total number of observed coincidence periods equal to or less than a given time interval is plotted against the time interval. The integral curve so obtained is exponential within the errors of the experiment and represents the decay of Em$^{218}$. The mean interval is 0.027~sec.\ corresponding to a half-life of 0.019~sec.'' This value is within a factor of two of the currently accepted half-life of 35(5)~ms.

\subsection*{$^{219}$Rn}
In the 1903 article ``Ueber den Emanationsk\"orper aus Pechblende und \"uber Radium'' Giesel identified a new emanation which was later identified as $^{219}$Rn \cite{1903Gie01}. The emanation was separated from a pitchblende sample. ``In den erw\"ahnten ca.\ 2~g m\"ussten also mindestens 2 Milligramm des fraglichen Elementes enthalten sein. Dass dasselbe nicht Radium oder Polonium sein kann, ist nach dem Gegebenen wohl ausgeschlossen. Von einer sonst aus practischen Gr\"unden \"ublichen Namengebung des hypothetischen Elementes sehe ich vorl\"aufig ab...'' [At least 2~mg of the element in question should be in the mentioned 2~g. Based on the presented facts it is probably ruled out that this substance can be radium or polonium. For now I refrain from the customary naming of the hypothetical element.] A month later Debierne independently observed the actinium emanation and reported that it disappear rapidly \cite{1903Deb01}. The half-life of $^{219}$Rn is 3.96(1)~s.

\subsection*{$^{220}$Rn}
Rutherford reported the observation of an activity from radium later identified as $^{220}$Rn in the 1900 article ``A radio-active substance emitted from thorium compounds'' \cite{1900Rut01}. Thorium oxide samples were used to study the activity of the ``emanation'': ``...I have found that thorium compounds continuously emit radio-active particles of some kind, which retain their radio-active powers for several minutes. This `emanation,' as it will be termed for shortness, has the power of ionizing the gas in its neighbourhood and of passing through thin layers of metals, and, with great ease, through considerable thicknesses of paper... The emanation passes through a plug of cotton-wool without any loss of its radio-active powers. It is also unaffected by bubbling through hot or cold water, weak or strong sulphuric acid. In this respect it acts like an ordinary gas... The result shows that the intensity of the radiation has fallen to one-half its value after an interval of about one minute.'' This half-life agrees with the currently accepted value of 55.6(1)~s.

\subsection*{$^{221}$Rn}
Momyer and Hyde reported the observation of $^{221}$Rn in the 1956 paper ``Properties of Em$^{221}$'' \cite{1956Mom01}. Thorium targets were bombarded with 110~MeV protons from the 184-inch Berkeley cyclotron. Alpha-decay spectra were measured following chemical separation. ``These results lead directly to the conclusion that a beta-emitting Em$^{221}$ with a 25-minute half-life is present in the samples and is giving rise to the known Fr$^{221}$ chain.'' The quoted value is the currently adopted half-life.

\subsection*{$^{222}$Rn}
In 1899 P. Curie and M. Curie reported the observation of an activity in radium samples later identified as $^{222}$Rn in ``Sur la radioactivit\'e provoqu\'ee par les rayons de Becquerel'' \cite{1899Cur01}. The radioactivity of polonium and radium samples was studied by measuring current due to the ionization of air. ``Si l'on soustrait la plaque activ\'ee a l'influence de la substance radioactive, elle reste radioactive pendant plusieurs jours. Toutefois, cette radioactivit\'e induite va en d\'ecroissant, d'abord tr\`es rapidement, ensuite de moins en moins vite et tend \'a dispara\^itre suivant une loi asymptotique.'' [Subtracting the contribution of the activated plate due to the radioactive substance, it remains radioactive for several days. However, the induced radioactivity is decreasing, first very rapidly, then slower and slower and tends to disappear asymptotically.] The currently accepted half-life of $^{222}$Ra is 3.8235(3)~d.

\subsection*{$^{223,224}$Rn}
Butement and Robinson announced the discovery of $^{223}$Rn and $^{224}$Rn in the 1964 paper ``New isotopes of emanation'' \cite{1964But01}. Thorium metal powder was irradiated with a 370~MeV proton beam from the Liverpool synchrocyclotron. The half-lives of $^{223}$Rn and $^{224}$Rn were determined by the milking technique, where the activities were measured with a ZnS-Ag alpha scintillation counter. ``The half-life of $^{224}$Em was obtained by extrapolating the decay curves of 3.6~day $^{224}$Ra to the time of milking, and plotting these extrapolated values against time of milking. The value obtained for the half life of $^{224}$Em is 114$\pm$6~min., the error being the standard deviation of the mean of six experiments... These experiments were very similar to those on $^{224}$Em, except that the intervals between milkings were shorter [because of the shorter half life of $^{223}$Em], and it was necessary to count the radium samples for some 50-60 days in order to follow the decay of 11.6~day $^{223}$Ra after 3.6~day $^{224}$Ra had decayed out.. The value obtained for the half life of $^{223}$Em is 43$\pm$5~min, where the error is the standard deviation on the mean of six experiments.'' The half-life of 43(5)~min for $^{223}$Rn is within a factor of two of the accepted value of 24.3(4)~min and the half-life of 114(6)~min for $^{224}$Rn agrees with currently accepted value of 107(3)~min.

\subsection*{$^{225,226}$Rn}
Hansen et al.\ reported the first observation of $^{225}$Rn and $^{226}$Rn in the paper ``Decay characteristics of short-lived radio-nuclides studied by on-line isotope separator techniques'' in 1969 \cite{1969Han01}. Protons of 600~MeV from the CERN synchrocyclotron bombarded a molten tin target and $^{225}$Rn and $^{226}$Rn were separated using the ISOLDE facility. The paper summarized the ISOLDE program and did not contain details about the individual nuclei but the results were presented in a table. The measured half-lives of 4.5(3)~min for $^{225}$Rn and 6.0(5)~min for $^{226}$Rn agree with the currently adopted values of 4.66(4)~min and 7.4(1)~min, respectively.

\subsection*{$^{227}$Rn}
In 1986 Borge et al.\ reported the observation of $^{227}$Rn in the article ``New isotope $^{227}$Rn and revised halflives for $^{223}$Rn and $^{226}$Rn'' \cite{1986Bor01}. ThC$_2$ was bombarded with 600~MeV protons from the CERN synchrocyclotron. Decay curves were measured with a 4$\pi$ plastic scintillation detector following mass separation with the ISOLDE on-line separator. ``These results yielded halflives of 23$\pm$1~s for the previously unknown isotope $^{227}$Rn and 2.52$\pm$0.05~min for $^{227}$Fr.'' This value is included in the calculation of the current half-life.

\subsection*{$^{228}$Rn}
$^{228}$Rn was first discovered by Borge et al.\ and the results were published in the 1989 paper ``The new neutron-rich isotope $^{228}$Rn'' \cite{1989Bor01}. The CERN synchrocyclotron was used to bombard a $^{232}$Th target with 600~MeV protons. Decay curves were measured with a 4$\pi$ plastic scintillation detector following mass separation with the ISOLDE II on-line separator. ``From the growth and decay pattern of the Ra K$_{\alpha1}$X-rays and the two strongest $\gamma$-lines from the decay of $^{228}$Fr at 141 and 474~keV a half-life of 36$\pm$2~s was obtained for $^{228}$Fr when the value of 65~s has been kept fixed for the precursor $^{228}$Rn, and a half-life of 62$\pm$3~s for $^{228}$Rn resulted when the value of 38~s has been kept fixed for the daughter nucleus $^{228}$Fr.'' The quoted half-life is the currently accepted value.

\subsection*{$^{229}$Rn}
Neidherr et al.\ announced the discovery of $^{229}$Rn in the 2009 article ``Discovery of $^{229}$Rn and the structure of the heaviest Rn and Ra isotopes from Penning-trap mass measurements'' \cite{2009Nei01}. A UC$_X$ target was bombarded with 1.4~GeV protons from the CERN proton synchroton booster accelerator. $^{229}$Rn was measured with the double Penning-trap mass spectrometer ISOLTRAP after mass separation with the on-line isotopes separator ISOLDE. ``This measurement gives a half-life of 12$^{+1.2}_{-1.3}$~s for a nuclide with mass number 229 then delivered to ISOLTRAP.'' The quoted value is the currently adopted half-life.

\subsection*{$^{230,231}$Rn}
$^{230}$Rn and $^{231}$Rn were discovered by Alvarez-Pol and the results were published in the 2010 paper ``Production of new neutron-rich isotopes of heavy elements in fragmentation reactions of $^{238}$U projectiles at 1A GeV'' \cite{2010Alv01}. A beryllium  target was bombarded with a 1~A GeV $^{238}$U beam from the GSI SIS synchrotron. The isotopes were separated and identified with the high-resolving-power magnetic spectrometer FRS. ``To search for new heavy neutron-rich nuclei, we tuned the FRS magnets for centering the nuclei $^{227}$At, $^{229}$At, $^{216}$Pb, $^{219}$Pb, and $^{210}$Au along its central trajectory. Combining the signals recorded in these settings of the FRS and using the analysis technique previously explained, we were able to identify 40 new neutron-rich nuclei with atomic numbers between Z=78 and Z=87; $^{205}$Pt, $^{207-210}$Au, $^{211-216}$Hg, $^{214-217}$Tl, $^{215-220}$Pb, $^{219-224}$Bi, $^{223-227}$Po, $^{225-229}$At, $^{230,231}$Rn, and $^{233}$Fr.''

%*********************************************************
%*********************************************************

\section{$^{199-233}$Fr}\vspace{0.0cm}

The element francium was discovered by Perey in 1939 by observing the $\beta$-decay of $^{223}$Fr in the natural actinium radioactive decay chain \cite{1939Per01}. Earlier incorrect observations of francium are described and referenced in the 2005 article ``Francium (atomic number 87), the last discovered natural element'' on the occasion of the 30$^{th}$ anniversary of Marguerite Perey's death in 1975 \cite{2005Adl01}. The name francium was officially accepted at the 15$^{th}$ IUPAC conference in Amsterdam in 1949 \cite{1949IUC01}. Perey had originally suggested the symbol Fa but agreed it to be changed to Fr \cite{2005Adl01}.

\begin{figure}
	\centering
	\includegraphics[scale=.7]{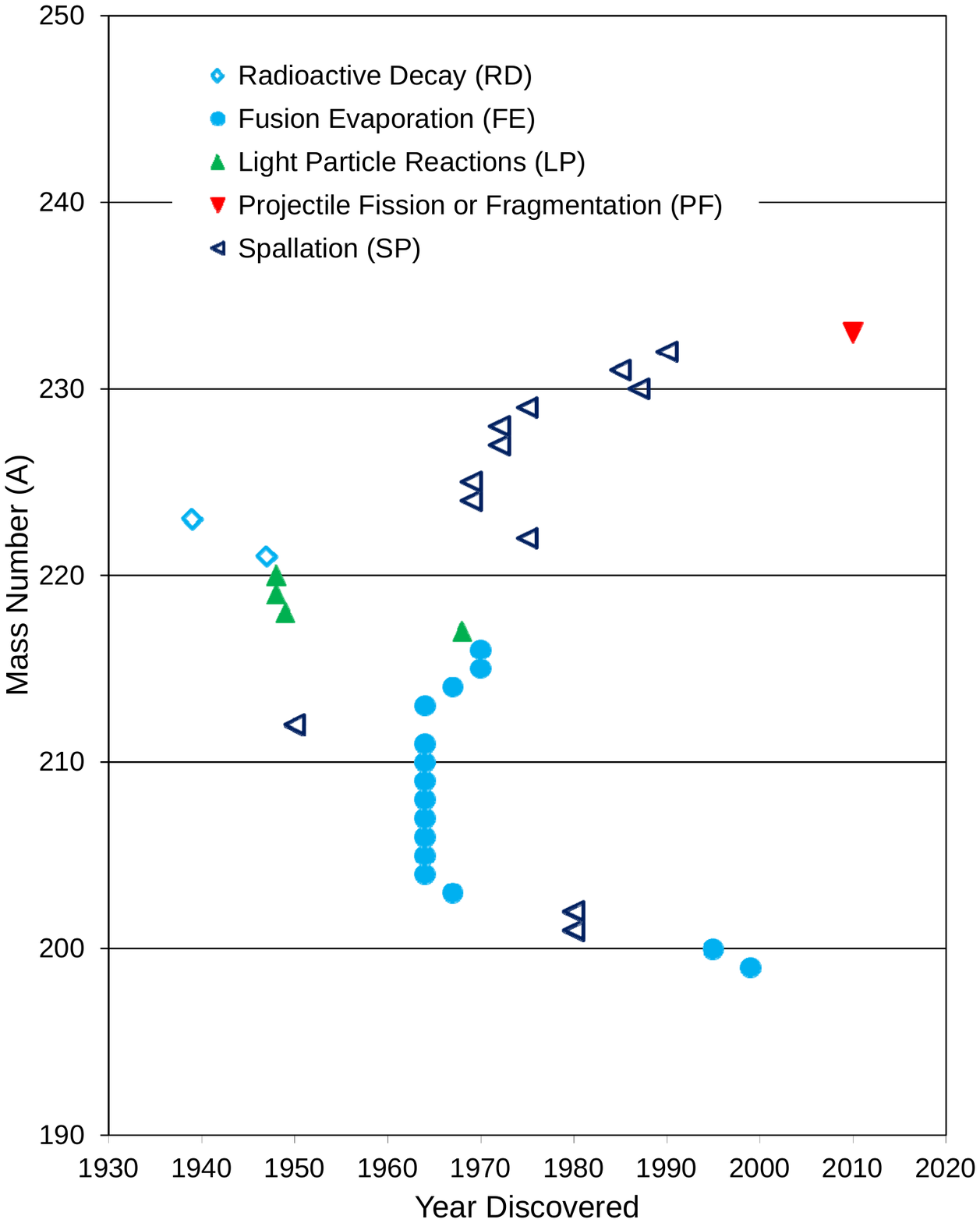}
	\caption{Francium isotopes as a function of time when they were discovered. The different production methods are indicated.}
\label{f:year-francium}
\end{figure}

Thirty-five francium isotopes from A = 193--231 have been discovered so far and according to the HFB-14 model \cite{2007Gor01} about 40 additional francium isotopes could exist. Figure \ref{f:year-francium} summarizes the year of first discovery for all francium isotopes identified by the method of discovery: radioactive decay (RD), fusion evaporation reactions (FE), light-particle reactions (LP), projectile fission or fragmentation (PF), and spallation (SP). In the following, the discovery of each francium isotope is discussed in detail and a summary is presented in Table 1.

\subsection*{$^{199}$Fr}
Tagaya et al.\ reported the discovery of $^{199}$Fr in the 1999 paper ``The $\alpha$-decay energies and halflives of $^{195g,m}$At and $^{199}$Fr'' \cite{1999Tag01}. $^{169}$Tm targets were bombarded with a 215~MeV $^{36}$Ar beam from the RIKEN ring cyclotron to form $^{199}$Fr in (6n) fusion-evaporation reactions. Recoils were separated with the gas-filled recoil separator GARIS and implanted in a position sensitive detector which also recorded subsequent $\alpha$ decay. ``The E$_\alpha$ and T$_{1/2}$ of $^{199}$Fr are 7655$\pm$40~keV and 12$^{+10}_{-4}$~ms, respectively.'' The quoted value is the currently accepted half-life.

\subsection*{$^{200}$Fr}
In the 1995 article ``New $\alpha$-decaying neutron deficient isotopes $^{197}$Rn and $^{200}$Fr,'' Morita et al.\ announced the identification of $^{200}$Fr \cite{1995Mor01}. A 186~MeV $^{36}$Ar beam from the RIKEN ring cyclotron bombarded a $^{169}$Tm target forming $^{200}$Fr in (5n) fusion-evaporation reactions. Reaction products were separated with the gas-filled recoil separator GARIS and implanted in a position-sensitive silicon detector which also measured subsequent $\alpha$ decay. ``The $\alpha$-decay energies (half-lives) of $^{197}$Rn, $^{197m}$Rn and $^{200}$Fr have been determined to be 7261$\pm$30~keV (51$^{+35}_{-15}$~ms), 7370$\pm$30~keV (18$^{+9}_{-5}$~ms), and 7500$\pm$30~keV, (570$^{+270}_{-140}$~ms), respectively.'' This value does not agree with the currently accepted value of 49(4)~ms. We credit Morita et al.\ with the discovery of $^{200}$Fr because they measured the correct decay energy and correlated the events with known properties of the daughter nucleus $^{196}$At. Three months later Enquist et al.\ \cite{1996Enq01} independently reported a half-life of 19$^{+13}_{-6}$~ms which also disagrees with the present value.

\subsection*{$^{201,202}$Fr}
The first observation of $^{201}$Fr and $^{202}$Fr was reported in ``Alpha decay studies of new neutron-deficient francium isotopes and their daughters'' by Ewan et al.\ \cite{1980Ewa01}. A uranium target was bombarded with 600~MeV protons from the CERN synchrocyclotron producing  $^{201}$Fr and $^{202}$Fr in spallation reactions. Alpha-particle spectra were measured with a silicon surface-barrier detector following mass separation with the isotope separator ISOLDE. ``The only hitherto unreported line in the spectrum is the 7388$\pm$15~keV line, whose decay, as obtained from the measurement with the position-sensitive detector, is shown in the inset of [the figure]. This line is assigned to $^{201}$Fr, for which a half-life of 48$\pm$15~ms thus was derived... The singles alpha spectrum observed from the decay of a source collected at mass 202 is shown in the lower part of [the figure]. In addition to previously known lines, mainly coming from heavier francium isotopes in analogy with the A=201 spectrum, a strong alpha line with an energy of 7251$\pm$10~keV is present... The new line is assigned to $^{202}$Fr, and the half-life was deduced to be 0.34$\pm$0.04~s.'' The measured half-lives of 48(15)~ms for $^{201}$Fr and 0.34(4)~s for $^{202}$Fr agree with the currently adopted values of 62(5)~ms and 0.30(5)~s, respectively.

\subsection*{$^{203}$Fr}
Valli et al.\ announced the discovery of $^{203}$Fr in the 1967 article ``Alpha decay of neutron-deficient francium isotopes'' \cite{1967Val02}. The Berkeley heavy ion linear accelerator was used to bombard $^{197}$Au and $^{205}$Tl targets with $^{16}$O and $^{12}$C beams with energies up to 166 and 126~MeV, respectively. Recoil products were collected on a catcher foil which was quickly positioned in front of a Si(Au) surface-barrier detector which measured subsequent $\alpha$ decay. ``Francium-203. The peak at 7.130~MeV with a half-life of 0.7$\pm$0.3~sec is visible only in the spectra taken at the highest beam energies. Comparison of excitation functions indicates that the peak belongs to a francium isotope lighter than $^{204}$Fr, most probably to $^{203}$Fr.'' This value is consistent with the currently adopted value of 0.30(5)~s.

\subsection*{$^{204-211}$Fr}
In 1964, Griffioen and MacFarlane reported the identification of $^{204}$Fr, $^{205}$Fr, $^{206}$Fr, $^{207}$Fr, $^{208}$Fr, $^{209}$Fr, $^{210}$Fr, and $^{211}$Fr in the paper ``Alpha-decay properties of some francium isotopes near the 126-neutron closed shell'' \cite{1964Gri01}. $^{197}$Au, $^{203,205}$Tl, and $^{208}$Pb targets were bombarded with $^{16}$O, $^{12}$C, and $^{11}$B beams with energies up to 10.38~MeV/amu from the Berkeley HILAC. Recoil products were collected on a catcher foil which was positioned in front of a gold surface-barrier detector which measured subsequent $\alpha$ decay. ``B. Fr$^{211}$ and Fr$^{210}$: ... These facts seem to indicate that this group is due to two different isotopes, Fr$^{211}$ and Fr$^{210}$, which were formed by (C$^{12}$,6n) and (C$^{12}$,7n) reactions, respectively... C. Fr$^{209}$ and Fr$^{208}$: ... Once again there were indications that this alpha group is a result of two different isotopes. The Tl$^{203}$+C$^{12}$ excitation function is somewhat broadened and distorted and no other alpha groups with an excitation function corresponding to a (C$^{12}$,7n) reaction were found. The excitation function from the Au$^{197}$+O$^{16}$ system is also consistent with the assignment to Fr$^{209}$ and Fr$^{208}$... D. Fr$^{207}$ and Fr$^{206}$: ... Since, as has been mentioned before and will be discussed later, the cross section for the Au$^{197}$(O$^{16}$,7n)Fr$^{206}$ reaction is probably about one fourth the value of the Au$^{197}$(O$^{16}$,6n)Fr$^{207}$ reaction cross section, we cannot see any great effect on the excitation function for this group. Several things have led us to the conclusion that this is both Fr$^{207}$ and Fr$^{206}$... E. Fr$^{205}$ and Fr$^{204}$: ... Since it follows the excitation function for the 6.91-MeV group, this would identify this group as the parent of At$^{201}$ namely, Fr$^{205}$. The assignment of the 7.02-MeV group to Fr$^{204}$ is based on the excitation-function data and on alpha decay systematics.'' The measured half-lives of 2.0(5)~s ($^{204}$Fr), 3.7(4)~s ($^{205}$Fr), 15.8(4)~s ($^{206}$Fr), 18.7(8)~s ($^{207}$Fr), 37.5(20)~s ($^{208}$Fr), 54.7(10)~s ($^{209}$Fr), 159(5)~s ($^{210}$Fr), and 186(4)~s ($^{211}$Fr) are close to the currently adopted values of 1.8(3)~s, 3.92(4)~s, 15.9(1)~s, 14.8(1)~s, 59.1~(s), 50.0(3)~s, 3.18(6)~min, and 3.10(2)~min, respectively.

\subsection*{$^{212}$Fr}
Hyde et al.\ reported the first observation of $^{212}$Fr in the 1950 paper ``Low mass francium and emanation isotopes of high alpha-stability'' \cite{1950Hyd01}. Thorium foils were bombarded with up to 350~MeV protons from the Berkeley 184-inch cyclotron. $^{212}$Fr was chemically separated and alpha spectra were measured with an ionization chamber. ``Fr$^{212}$, with an apparent half-life of 19.3 minutes for branching decay by alpha-emission (44 percent) to At$^{208}$ and by orbital electron-capture (56 percent) to Em$^{212}$, has been found.'' This half-life is included in the calculation of the currently adopted value.

\subsection*{$^{213}$Fr}
In 1964 Griffioen and MacFarlane reported the identification of $^{213}$Fr in the paper ``Alpha-decay properties of some francium isotopes near the 126-neutron closed shell'' \cite{1964Gri01}. $^{197}$Au, $^{203,205}$Tl, and $^{208}$Pb targets were bombarded with $^{16}$O, $^{12}$C, and $^{11}$B beams with energies up to 10.38~MeV/amu from the Berkeley HILAC. Recoil products were collected on a catcher foil which was positioned in front of a gold surface-barrier detector which measured subsequent $\alpha$ decay. ``A. Fr$^{213}$: [The figure] shows an alpha-particle spectrum of the activity collected while bombarding Tl$^{205}$ with 86-MeV C$^{12}$ ions. A strong group is seen at 6.77$\pm$0.01~MeV alpha particle energy. This activity decays with a half-life of 33.7$\pm$1.5~sec.'' This value is included in the calculation of the currently accepted value.

\subsection*{$^{214}$Fr}
Rotter et al.\ observed $^{214}$Fr in 1967 and reported their results in the paper ``The new isotope Ac$^{216}$'' \cite{1967Rot01}. A 78~MeV $^{12}$C beam from the Dubna 1.5~m cyclotron bombarded a bismuth target forming actinium in (xn) reactions. $^{214}$Fr was populated by $\alpha$ decay of $^{218}$Ac. Recoil nuclei were collected on an aluminum foil and $\alpha$-particle spectra were measured with a silicon surface barrier detector. ``We obtained the following $\alpha$-particle energies: Rn$^{213}$ - 8.14~MeV, Fr$^{214}$ - 8.53~MeV, and Ra$^{215}$ - 8.73~MeV.'' Rotter et al.\ did not consider this observation a new discovery referring to a conference abstract \cite{1962Gri01}. The observation corresponds to an isomeric state with a currently accepted half-life of 3.35(5)~ms.

\subsection*{$^{215,216}$Fr}
In the 1970 article ``Production and decay properties of protactinium isotopes of mass 222 to 225 formed in heavy-ion reactions,'' Borggreen et al.\ identified $^{215}$Fr and $^{216}$Fr \cite{1970Bor01}. The Berkeley heavy-ion linear accelerator (HILAC) was used to bombard $^{208}$Pb and $^{205}$Tl targets with $^{19}$F and $^{22}$Ne beams forming $^{224}$Pa and $^{223}$Pa in (3n) and (4n) fusion-evaporation reactions, respectively. $^{216}$Fr and $^{215}$Fr were then populated by subsequent $\alpha$-decay. Recoil products were deposited by a helium gas stream on a metal surface located in front of a gold surface-barrier detector which recorded the subsequent $\alpha$ decay. ``Francium-216 appears to emit a single $\alpha$ group of 9.005$\pm$0.010~MeV which fits smoothly on the francium curve in [the figure]... The time-parameter information associated with the data sorting displayed in [the figure] yielded a 0.70$\pm$0.02~$\mu$sec half-life for $^{216}$Fr... The assignment of the 9.365-MeV group to $^{215}$Fr seems particularly secure owing to the very restricted number of possible assignments of $\alpha$ groups above 9.3-MeV energy.'' For $^{215}$Fr only an upper limit of $<$500~ns was given. The currently accepted value is 86(5)~ns. The measured half-life for $^{216}$Fr is the presently adopted value.

\subsection*{$^{217}$Fr}
In the 1968 article ``New neptunium isotopes, $^{230}$Np and $^{229}$Np'' Hahn et al.\ reported the observation of $^{217}$Fr \cite{1968Hah01}. Enriched $^{233}$U targets were bombarded with 32$-$41.6~MeV protons from the Oak Ridge Isochronous Cyclotron forming $^{229}$Np in (p,5n) reactions. Reaction products were implanted on a catcher foil which was periodically rotated in front of a surface barrier Si(Au) detector. Isotopes populated by subsequent $\alpha$ emission were measured. ``The $\alpha$-particle energies found for the $^{225}$Pa series are more precise than the previously available values: $^{225}$Pa, 7.25$\pm$0.02~MeV (new value); $^{221}$Ac, 7.63$\pm$0.02~MeV; $^{217}$Fr, 8.31$\pm$0.02~MeV and $^{213}$At, 9.06$\pm$0.02~MeV.'' The observation of $^{217}$Fr was not considered new, referring to an unpublished thesis \cite{1951Key01}.

\subsection*{$^{218}$Fr}
Meinke et al.\ reported the observation of $^{218}$Fr in the 1949 paper ``Three additional collateral alpha-decay chains'' \cite{1949Mei01}. Thorium was bombarded with 150~MeV deuterons from the Berkeley 184-inch cyclotron. The $\alpha$-decay chain from $^{226}$Pa was measured following chemical separation. ``General considerations with regard to the method of formation and half-life of the parent substance, and the energies of all the members of the series suggest a collateral branch of the 4n+2 family: $_{91}$Pa$^{226}\overset{\alpha}{\rightarrow}_{89}$Ac$^{222}\overset{\alpha}{\rightarrow}_{87}$Fr$^{218}\overset{\alpha}{\rightarrow}_{85}$At$^{214}\overset{\alpha}{\rightarrow}_{85}$Bi$^{210}$(RaE).'' In a table summarizing the energies and half-lives of the decay chain only the $\alpha$-decay energy was given for $^{216}$Rn stating a calculated half-life of about 10$^{-2}$~s. The currently accepted half-life is 1.0(6)~ms.

\subsection*{$^{219,220}$Fr}
In ``Artificial collateral chains to the thorium and actinium families,'' Ghiorso et al.\ discovered $^{219}$Fr and $^{220}$Fr in 1948 \cite{1948Ghi01}. Thorium targets were irradiated with 80~MeV deuterons from the Berkeley 184-inch cyclotron. The $\alpha$-decay chains beginning at $^{227}$Pa and $^{228}$Pa were measured following chemical separation. ``Prominent soon after bombardment are a number of alpha-particle groups, which decay with the 38-minute half-life of the protactinium parent. These are due to the following collateral branch of the 4n+3 radioactive family: $_{91}$Pa$^{227}\stackrel{\alpha}{\longrightarrow}_{89}$Ac$^{223}\stackrel{\alpha}{\longrightarrow}_{87}$Fr$^{219}\stackrel{\alpha}{\longrightarrow}_{85}$At$^{215}\stackrel{\alpha}{\longrightarrow}$... After the decay of the above-described series, a second group of alpha-particle emitters can be resolved. This second series, which decays with the 22-hour half-life of its protactinium parent, is a collateral branch of the 4n radioactive family as follows: $_{91}$Pa$^{228}\stackrel{\alpha}{\longrightarrow}_{89}$Ac$^{224}\stackrel{\alpha}{\longrightarrow}_{87}$Fr$^{220}\stackrel{\alpha}{\longrightarrow}_{85}$At$^{216}\stackrel{\alpha}{\longrightarrow}$...'' The decay energies and half-lives of the decay chains were listed in a table, assigning half-lives of $\sim$10$^{-4}$~s and $\sim$30~s to $^{219}$Fr and $^{220}$Fr, respectively. The currently adopted half-lives for $^{219}$Fr and $^{220}$Fr are 20(2)~ms and 27.4(3)~s, respectively.

\subsection*{$^{221}$Fr}
Hagemann et al.\ discovered $^{221}$Fr in 1947 in ``The (4n+1) radioactive series: the decay products of U$^{233}$'' \cite{1947Hag01}. The half-lives and $\alpha$- and $\beta$-decay energies of the nuclides in the decay chain of $^{233}$U were measured. ``These decay products, which constitute a substantial fraction of the entire missing, 4n+1, radioactive series are listed together with their radioactive properties, in [the table].'' The measured half-life of 4.8~min agrees with the presently accepted value of 4.9(2)~min. Hagemann et al.\ acknowledge the simultaneous observation of $^{221}$Fr by English et al.\ which was submitted only a day later and published in the same issue of Physical Review on the next page \cite{1947Eng01}.

\subsection*{$^{222}$Fr}
Westgaard et al.\ identified $^{222}$Fr in the 1975 paper ``Beta-decay energies and masses of short-lived isotopes of rubidium, caesium, francium, and radium'' \cite{1975Wes01}. Lanthanum, yttrium-lanthanum, and thorium-lanthanum targets were irradiated with 600~MeV protons from the CERN synchrocyclotron. Beta- and gamma-rays were measured following mass separation with the ISOLDE on-line separator at CERN. ``The decay of 15~min $^{222}$Fr: ... The singles $\beta$ spectrum measured in our experiment showed a flat tail of low intensity, extending to much higher energies than the main portion of the data. After subtraction of this tail, presumably due to $\alpha$ particles from $^{222}$Ra, a FK analysis gave for the endpoint energy E$^{max}_\beta$=1.78$\pm$0.02~MeV.'' They measured half-life of 14.8~min agrees with the currently adopted value of 14.2(3)~min.

\subsection*{$^{223}$Fr}
Perey discovered $^{223}$Fr in 1939 as reported in ``Sur un \'el\'ement 87, d\'eriv\'e de l'actinium'' \cite{1939Per01}. $^{223}$Fr was observed within the natural actinium radioactive decay chain and populated by $\alpha$ decay from $^{227}$Ac. Beta-decay curves were measured following chemical separation. ``En ajoutant du chlorure de c\ae sium \`a l'eau m\`ere et en pr\'ecipitant par une solution de perchlorate de sodium, il se forme des cristaux qui entra\^inent l'activit\'e: celle-ci d'ecro\^it exponentiellement avec la p\'eriode de 21 minutes $\pm$ 1. [By adding liquid cesium chloride and precipitating a solution of sodium perchlorate crystals are formed that cause an activity which decreases exponentially with the period of 21 minutes $\pm$ 1.] This half-life agrees with the presently adopted value of 22.00(7)~min. This observation of $^{223}$Fr also represented the discovery of the element francium.

\subsection*{$^{224-226}$Fr}
Hansen et al.\ reported the first observation of $^{224}$Fr, $^{225}$Fr and $^{226}$Fr in the paper ``Decay characteristics of short-lived radio-nuclides studied by on-line isotope separator techniques'' in 1969 \cite{1969Han01}. Protons of 600~MeV from the CERN synchrocyclotron bombarded a molten tin target and $^{224}$Fr, $^{225}$Fr and $^{226}$Fr were separated using the ISOLDE facility. The paper summarized the ISOLDE program and did not contain details about the individual nuclei but the results were presented in a table. The measured half-life of 2.67(20)~min for $^{224}$Fr agrees with the presently adopted value of 3.33(10)~min and the 3.9(2)~min for $^{225}$Fr is included in the calculation of the currently accepted half-life of 3.95(14)~min. The half-life of 1.43(23)~min for $^{226}$Fr is within a factor of two of the present value of 49(1)~s.

\subsection*{$^{227,228}$Fr}
In 1972 Klapisch et al.\ reported the first observation of $^{227}$Fr and $^{228}$Fr in ``Half-life of the new isotope $^{32}$Na; Observation of $^{33}$Na and other new isotopes produced in the reaction of high-energy protons on U'' \cite{1972Kla01}. Uranium targets were bombarded with 24 GeV protons from the CERN proton synchrotron. $^{227}$Fr and $^{228}$Fr were identified by on-line mass spectrometry and decay curves were measured. ``Following the same procedure as for Na, the isotopes $^{48}$K, $^{49}$K, and $^{50}$K were found. However, their half-lives were not short compared with the diffusion time, and hence could not be determined. We also observed the new neutron-rich isotopes $^{227}$Fr and $^{228}$Fr produced in the spallation of the uranium target.'' The presently accepted half-lives are 2.47(3)~min and 38(1)~s for $^{227}$Fr and $^{228}$Fr, respectively.

\subsection*{$^{229}$Fr}
In 1975 the discovery of $^{229}$Fr by Ravn et al.\ was announced in the paper ``Short-lived isotopes of alkali and alkaline-earth elements studied by on-line isotope separator techniques'' \cite{1975Rav01}. A thorium plus lanthanum target was bombarded with protons from the CERN synchrocyclotron. Beta-ray decay curves were measured with a 4$\pi$ plastic detector following mass separation with the isotope separator ISOLDE. ``The following half-lives of new nucleides have been determined: ... $^{229}$Fr (50$\pm$20)~sec.'' This half-life agrees with the presently adopted value of 50.2(20)~s.

\subsection*{$^{230}$Fr}
In the 1987 article ``Collective states in $^{230}$Ra fed by $\beta^-$ decay of $^{230}$Fr,'' Kurcewicz et al.\ identified $^{230}$Fr \cite{1987Kur01}. Francium was produced by spallation of $^{238}$U with 600~MeV protons from the CERN synchrocyclotron. Gamma-ray singles and $\gamma-\gamma$ coincidences were measured with Ge(Li) detectors after mass separation with the on-line separator ISOLDE II. ``A half-life of 19.1$\pm$0.5~s for $^{230}$Fr has been obtained by means of multispectra analysis using cycles of 20~s collection time followed by 6$\times$7~s measuring time.'' This value is the currently accepted half-life.

\subsection*{$^{231}$Fr}
The discovery of $^{231}$Fr was reported in the 1985 paper ``The new neutron-rich nuclei $^{231}$Fr and $^{231}$Ra'' by Hill et al.\ \cite{1985Hil01}. Francium was produced by spallation of $^{238}$U with 600~MeV protons from the CERN synchrocyclotron. Beta-particles and $\gamma$-rays were measured with a plastic scintillator and two Ge(Li) detectors, respectively, following mass separation with the on-line separator ISOLDE II. ``With three other $\gamma$-lines, which are assigned to the $^{231}$Fr decay due to their half-lives, a weighted average of 17.5(8)~s is obtained for the half-life of $^{231}$Fr.'' The quoted value is included in the calculation of the currently adopted half-life.

\subsection*{$^{232}$Fr}
Mezlev et al.\ reported the discovery of $^{232}$Fr in the 1990 paper ``Search for delayed fission in neutron-rich nuclei'' \cite{1990Mez01}. A uranium target was bombarded with 1~GeV protons. Beta-, gamma-, and X-rays were measured with solid state detectors following mass separation with the on-line mass separator IRIS. ``Due to this technique the new isotopes $^{232}$Fr (T$_{1/2}$=5$\pm$1~s), $^{233}$Ra(T$_{1/2}$=30$\pm$5~s) and $^{234}$Ra (T$_{1/2}$=30$\pm$10~s) have been identified using the solid state detectors for the registration of the beta-, gamma-, X-radiation.'' The measured of 5(1)~s half-life for $^{232}$Fr agrees with the currently adopted value of 5.5(6)~s.

\subsection*{$^{233}$Fr}
$^{233}$Fr was discovered by Alvarez-Pol and the results were published in the 2010 paper ``Production of new neutron-rich isotopes of heavy elements in fragmentation reactions of $^{238}$U projectiles at 1A GeV'' \cite{2010Alv01}. A beryllium  target was bombarded with a 1~A GeV $^{238}$U beam from the GSI SIS synchrotron. The isotopes were separated and identified with the high-resolving-power magnetic spectrometer FRS. ``To search for new heavy neutron-rich nuclei, we tuned the FRS magnets for centering the nuclei $^{227}$At, $^{229}$At, $^{216}$Pb, $^{219}$Pb, and $^{210}$Au along its central trajectory. Combining the signals recorded in these settings of the FRS and using the analysis technique previously explained, we were able to identify 40 new neutron-rich nuclei with atomic numbers between Z=78 and Z=87; $^{205}$Pt, $^{207-210}$Au, $^{211-216}$Hg, $^{214-217}$Tl, $^{215-220}$Pb, $^{219-224}$Bi, $^{223-227}$Po, $^{225-229}$At, $^{230,231}$Rn, and $^{233}$Fr.''

%*********************************************************
%*********************************************************

\section{$^{201-234}$Ra}\vspace{0.0cm}

\begin{figure}
	\centering
	\includegraphics[scale=.7]{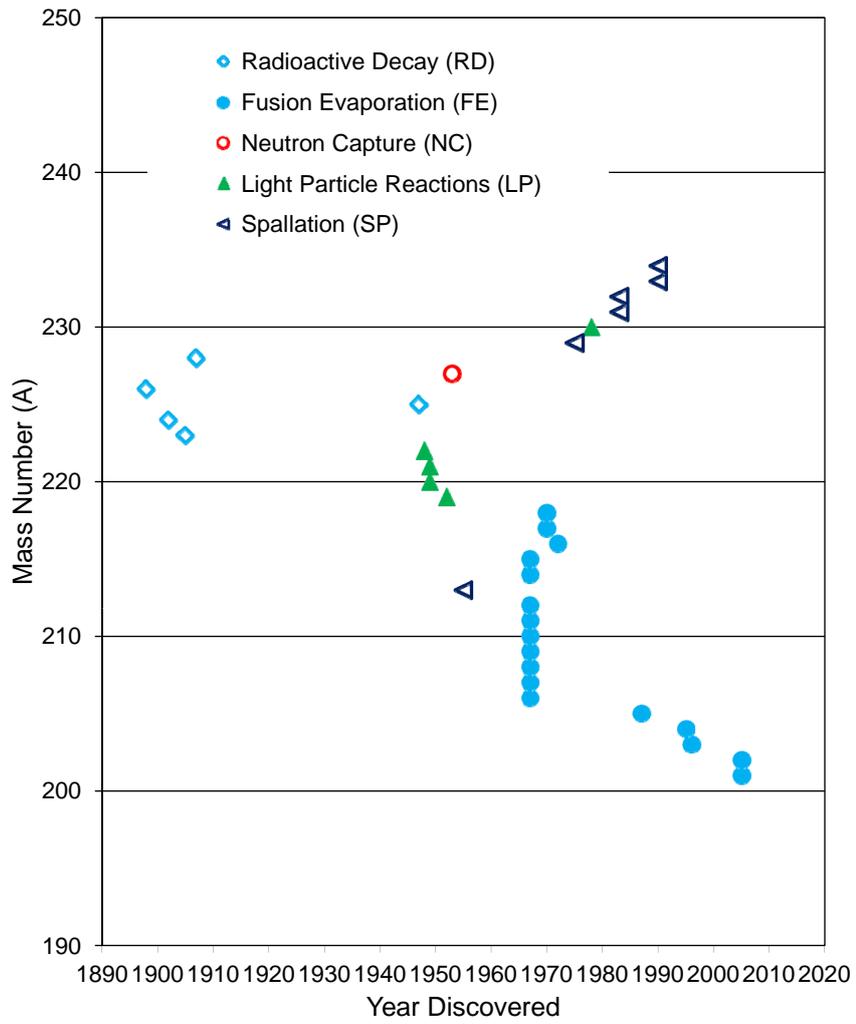}
	\caption{Radium isotopes as a function of time when they were discovered. The different production methods are indicated.}
\label{f:year-radium}
\end{figure}

The element radium was discovered in 1898 by P. Curie, M. Curie and G. B\'emont \cite{1898Cur02}. Thirty-four radium isotopes from A = 201--234 have been discovered so far and according to the HFB-14 model \cite{2007Gor01} about 50 additional radon isotopes could exist. Figure \ref{f:year-radium} summarizes the year of first discovery for all radon isotopes identified by the method of discovery: radioactive decay (RD), fusion evaporation reactions (FE), light-particle reactions (LP), neutron-capture reactions (NC), and spallation (SP). In the following, the discovery of each radon isotope is discussed in detail and a summary is presented in Table 1.

\subsection*{$^{201,202}$Ra}
$^{201}$Ra and $^{202}$Ra were first observed by Uusitalo et al.\ and the results were published in the 2005 paper ``$\alpha$ decay studies of very neutron-deficient francium and radium isotopes'' \cite{2005Uus01}. A $^{141}$Pr target was bombarded with 278-288~MeV $^{63}$Cu beams from the Jyv\"askyl\"a K-130 cyclotron forming $^{201}$Ra and $^{202}$Ra in (3n) and (2n) fusion-evaporation reactions, respectively. Reaction products were separated with the gas-filled recoil separator RITU and implanted in a position-sensitive silicon detector which measured subsequent $\alpha$ decay. ``We associate these measured values with the known activities of $^{197}$Rn$^m$ (from the 13/2$^+$ isomeric state) with E$_\alpha$ = 7356(7)~keV and T$_{1/2}$ = 19$^{+8}_{-4}$~ms and $^{193}$Po$^m$ (from the 13/2$^+$ isomeric state) with E$_\alpha$ = 7004(5)~keV and T$_{1/2}$ = 240(10)~ms [13], and thus the activity with E$_\alpha$ = 7905~keV and T$_{lt}$ = 2~ms can be identified to originate from a new even-odd radium isotope $^{201}$Ra... Thus the activity with E$_\alpha$ = 7740~keV and T$_{lt}$ = 46~ms can be identified to originate from the (0$^+$) ground state of a new even-even isotope $^{202}$Ra.'' The measured half-lives of 1.6$^{+7.7}_{-0.7}$~ms and 16$^{+30}_{-7}$~ms are the current values for $^{201}$Ra and $^{202}$Ra, respectively. A previously measured half-life of 0.7$^{+3.3}_{-0.3}$~ms for $^{202}$Ra \cite{1996Lei01} was evidently incorrect.

\subsection*{$^{203}$Ra}
In the 1996 paper ``Alpha decay studies of neutron-deficient radium isotopes,'' Leino et al.\ described the observation of $^{203}$Ra \cite{1996Lei01}. A $^{175}$Lu target was bombarded with 191$-$208~MeV $^{35}$Cl beams from the Jyv\"askyl\"a K130 cyclotron to produce $^{203}$Ra in the (7n) fusion-evaporation reaction. Reaction products were separated with the gas-filled recoil separator RITU and implanted in a position-sensitive PIPS detector which measured subsequent $\alpha$ decay. ``The alpha particle energy E$_\alpha$ and the half-life T$_{1/2}$ of an isomeric state of the new isotope, $^{203m}$Ra, have been determined to be (7615$\pm$20)~keV and (33$^{+22}_{-10}$)~ms, respectively. An assignment of another decay with E$_\alpha$ = (7577$\pm$20)~keV and T$_{1/2}$ = (1.1$^{+5.0}_{-0.5}$)~ms to $^{203g}$Ra is made on the basis of one observed three-decay chain.'' While the half-life for the ground state is not consistent with the currently adopted value of 31$^{+17}_{-9}$~ms the half-life for the isomeric state agrees with the present value of 24$^{+6}_{-4}$~ms.

\subsection*{$^{204}$Ra}
Leddy et al.\ reported the observation of $^{204}$Ra in the 1995 article ``$\alpha$ decay of a new isotope, $^{204}$Ra'' \cite{1995Led01}. $^{28}$Si beams with energies of 164 and 170~MeV from the Argonne Tandem-Linac Accelerator System bombarded a $^{182}$W target producing $^{204}$Ra in the (6n) fusion-evaporation reaction. Recoil products were separated with the Fragment Mass Analyzer (FMA) and implanted in a double-sided silicon strip detector which measured subsequent $\alpha$ decay. ``This assignment, summarized in [the table], constitutes the first observation of $\alpha$ decay from the ground state of $^{204}$Ra with an $\alpha$ energy of 7.488(12)~MeV and a half-life of 45$^{+55}_{-21}$~ms.'' The quoted half-life agrees with the currently accepted value of 57($^{+11}_{-5}$)~ms.

\subsection*{$^{205}$Ra}
$^{205}$Ra was first observed by He\ss berger et al.\ as reported in the 1987 paper ``Observation of two new alpha emitters with Z = 88'' \cite{1987Hes01}. A $^{159}$Tb target was irradiated with $^{51}$V beams from the GSI UNILAC. Recoils were separated by the velocity filter SHIP and implanted in an array of seven position-sensitive surface-barrier detectors. ``For $^{205}$Ra an $\alpha$ energy of E$_\alpha$=(7360$\pm$20)~keV and a half-life of T$_{1/2}$=(220$\pm$60)~ms were obtained.'' This value agrees with the currently adopted half-life of 210($^{+60}_{-40}$)~ms.

\subsection*{$^{206-212}$Ra}
In the 1967 paper ``On-line alpha spectroscopy of neutron-deficient radium isotopes,'' Valli et al.\ described the observation of $^{206}$Ra, $^{207}$Ra, $^{208}$Ra, $^{209}$Ra, $^{210}$Ra, $^{211}$Ra, and $^{212}$Ra \cite{1967Val03}. $^{197}$Au and $^{206}$Pb targets were bombarded with $^{19}$F and $^{12}$C beams from the Berkeley HILAC to produce $^{216}$Ra and $^{218}$Ra compound nuclei. Recoil products were deposited on a collector foil which was then placed in front of a Si(Au) surface barrier detector. ``C. Radium-212 and Radium-211: ...In careful measurements made at beam energies where $^{212}$Ra was predominant over $^{211}$Ra we determined an $\alpha$ energy of 6.896~MeV and a half-life of 13$\pm$2 sec. At beam energies where $^{211}$Ra was the principal activity we determined values of 6.910~MeV and 15$\pm$2~sec... D. Radium-210 and Radium-209: ...From measurements made at a 105-MeV $^{19}$F beam energy where $^{210}$Ra predominates over $^{209}$Ra we measured an $\alpha$ energy of 7.018~MeV and a half-life of 3.8$\pm$0.2~sec. Properties of $^{209}$Ra were measured on samples prepared at a beam energy of 140~MeV. The $\alpha$ energy and half-life are 7.008~MeV and 4.7$\pm$0.2~sec, respectively. E. Radium-208 and Radium-207: ...The half-life of the group was also determined at several beam energies. Values between 1.1 and 1.4~sec were observed, the shorter ones coming systematically from measurements at lower beam energies. Therefore, 1.3$\pm$0.2~sec is reported for $^{207}$Ra and 1.2$\pm$0.2~sec for $^{208}$Ra... F. Radium-206: The weak $\alpha$ group at 7.270~MeV in [the figure] belongs to $^{206}$Ra. Its half-life was measured to be 0.4$\pm$0.2~sec.'' These half-lives of 13(2)~s, 15(2)~s, 3.8(2)~s, 4.7(2)~s, 1.2(2)~s, 1.3(2)~s, and 0.4(2)~s for $^{212}$Ra, $^{211}$Ra, $^{210}$Ra, $^{209}$Ra, $^{208}$Ra, $^{207}$Ra, and $^{206}$Ra, respectively agree with the currently adopted values of 13.0(2)~s, 13(2)~s, 3.7(2)~s, 4.6(2)~s, 1.3(2)~s, 1.35$^{+0.22}_{-0.13}$~s, and 0.24(2)~s.

\subsection*{$^{213}$Ra}
Momyer and Hyde reported the observation of $^{213}$Rn in the 1955 article ``The influence of the 126-neutron shell on the alpha-decay properties of the isotopes of emanation, francium, and radium'' \cite{1955Mom01}. A lead target was bombarded with a $^{12}$C beam from the Berkeley 60-inch cyclotron to form $^{213}$Ra in the $^{206}$Pb($^{12}$C,5n) fusion-evaporation reaction. Alpha-particle spectra and decay curves were recorded with an $\alpha$ pulse analyzer following chemical separation. ``The half-life of the activity was 2.7$\pm$0.3~minutes, and the energy of the alpha particle was 6.90$\pm$0.04~MeV. After decay of the short-lived activity, several counts per minute of Em$^{209}$ were observed on the plate.'' This half-life agrees with the currently adopted value of 2.73(5)~min. Earlier Momyer et al.\ presented indirect evidence for the production of a $\sim$2~min half-life for $^{213}$Ra \cite{1952Mom01}.

\subsection*{$^{214,215}$Ra}
Rotter et al.\ observed $^{214}$Ra and $^{215}$Ra in 1967 and reported their results in the paper ``The new isotope Ac$^{216}$'' \cite{1967Rot01}. An 78~MeV $^{12}$C beam from the Dubna 1.5~m cyclotron bombarded a lead target forming radium in (xn) reactions. Recoil nuclei were collected on an aluminum foil and $\alpha$-particle spectra were measured with a silicon surface barrier detector. ``...an accuracy of 0.02-0.03~MeV is possible when determining $\alpha$-line energies if the absolute energy scale is based on an `internal' reference line. The latter was the 6.77~MeV line of Fr$^{213}$ in our experiments with bismuth bombardment; in the case of a lead target it was the 7.17~MeV line of Ra$^{214}$. These lines were clearly discriminated in all the measured spectra... We obtained the following $\alpha$-particle energies: Rn$^{213}$ - 8.14~MeV, Fr$^{214}$ - 8.53~MeV, and Ra$^{215}$ - 8.73~MeV.'' Rotter et al.\ did not consider these observations new discoveries referring to an earlier conference abstract \cite{1962Gri01} and a book by Hyde et al.\ \cite{1964Hyd01} which most probably also referred to the same conference abstract. Over a year later Valli et al.\ independently reported a 7.136(5)~MeV $\alpha$ energy for $^{214}$Ra \cite{1967Val03}. The presently accepted half-lives are 2.46(3)~s and 1.55(7)~ms for $^{214}$Ra and $^{215}$Ra, respectively.

\subsection*{$^{216}$Ra}
In the article ``In-beam alpha spectroscopy of N=128 isotones. Lifetimes of $^{216}$Ra and a new isotope $^{217}$Ac,'' Nomura et al.\ reported the observation of $^{216}$Ra in 1972 \cite{1972Nom01}. A $^{206}$Pb target was bombarded with 65$-$85~MeV $^{12}$C beams from the RIKEN IPCR cyclotron forming $^{216}$Ra in (4n) fusion-evaporation reactions. Alpha-particle spectra and decay curves were measured with a surface barrier Si detector. ``Time distributions of the ground-state decay of $^{216}$Ra and $^{217}$Ac are shown in [the figure], from which half-lives of $^{216}$Ra and $^{217}$Ac have been determined of 0.18$\pm$0.03$~\mu$s and 0.10$\pm$0.01~$\mu$s, respectively.'' The quoted value for $^{216}$Ra agrees with the currently adopted half-life of 182(10)~ns.

\subsection*{$^{217,218}$Ra}
In 1970, Torgerson and MacFarlane reported the first observation of $^{217}$Ra and $^{218}$Ra in ``Alpha decay of the $^{221}$Th and $^{222}$Th decay chains'' \cite{1970Tor01}. A 10.6~MeV/nucleon $^{16}$O beam from the Yale heavy ion accelerator was used to bombard a $^{208}$Pb target forming $^{221}$Th and  $^{222}$Th in (3n) and (2n) fusion-evaporation reactions, respectively. $^{217}$Ra and $^{218}$Ra were then populated by subsequent $\alpha$ decay. Recoil products were transported to a stainless steel surface with a helium jet and $\alpha$ spectra were measured with a Si(Au) surface barrier detector. ``$^{217}$Ra: A direct measurement of the half-life was made by gating the two prominent $^{221}$Th groups (8.146 and 8.472~MeV) and the $^{217}$Ra group to produce start and stop pulses respectively on a time-to-amplitude converter. The output of the converter was then routed into a pulse-height analyser, and the decay curve was obtained. Using this procedure, we have measured the half-life of $^{217}$Ra to be 4$\pm$2~$\mu$sec... $^{218}$Ra: ...We observed a group at 8.392~MeV with an intensity relative to the $^{222}$Th group that was expected if it were an $\alpha$-decay daughter of that nucleide... No half-life measurement could be made because of the weak intensity of this group.'' Only three days later Valli et al.\ submitted their measurements of 8.995(10)~MeV and 8.385(1)~MeV $\alpha$ energies for $^{217}$Ra and $^{218}$Ra, respectively \cite{1970Val01}. The measured half-life for $^{217}$Ra of 4(2)~$\mu$s is close to the presently accepted value of 1.63(17)~$\mu$s and the current half-life for $^{218}$Ra is 25.2(3)~$mu$s.

\subsection*{$^{219}$Ra}
In 1952, $^{219}$Ra was discovered by Meinke et al.\ and the results were reported in the paper ``Further work on heavy collateral radioactive chains'' \cite{1952Mei01}. Thorium nitrate targets were irradiated with a $^4$He beam from the Berkeley 184-inch cyclotron. $^{227}$U was chemically separated and the decay and energy of $\alpha$-particles were measured with an alpha-particle pulse analyzer. ``An additional short-lived chain collateral to the actinium (4n+3) natural radioactive family has also been partially identified. This chain decays as follows: U$^{227}\rightarrow$Th$^{223}\rightarrow$Ra$^{219}\rightarrow$Em$^{215}\rightarrow$Po$^{211}\rightarrow$Pb$^{207}$.'' An $\alpha$ energy of 8.0(1)~MeV was assigned to $^{219}$Ra. The presently accepted half-life is 10(3)~ms.

\subsection*{$^{220,221}$Ra}
Meinke et al.\ reported the observation of $^{220}$Ra and $^{221}$Ra in the 1949 paper ``Three additional collateral alpha-decay chains'' \cite{1949Mei01}. Thorium was bombarded with 100$-$120~MeV $^4$He beams from the Berkeley 184-inch cyclotron. Alpha-decay chains from $^{228}$U and $^{229}$U were measured following chemical separation. ``The irradiation of thorium with 100-Mev helium ions resulted in the observation of the following collateral branch of the artificial 4n$+$1, neptunium, radioactive family shown with Po$^{213}$ and its decay products: $_{92}$U$^{229}\overset{\alpha}{\rightarrow}_{90}$Th$^{225}\overset{\alpha}{\rightarrow}_{88}$Ra$^{221}\overset{\alpha}{\rightarrow}_{86}$Em$^{217}\ldots$ Immediately after 120-Mev helium ion bombardment of thorium the uranium fraction contains another series of five alpha-emitters, which is apparently a collateral branch of the 4n family: $_{92}$U$^{228}\overset{\alpha}{\rightarrow}_{90}$Th$^{224}\overset{\alpha}{\rightarrow}_{88}$Ra$^{220}\overset{\alpha}{\rightarrow}_{86}$Em$^{216}\ldots$'' In a table summarizing the energies and half-lives of the decay chains only the $\alpha$-decay energy was given for $^{220}$Ra stating a calculated half-life of 10$^{-2}$~s. The currently accepted half-life is 18(2)~ms. The measured half-life of 31.0(15)~s for $^{221}$Ra agrees with the presently adopted value of 28(2)~s.

\subsection*{$^{222}$Ra}
Studier and Hyde announced the discovery of $^{222}$Ra in the 1948 paper ``A new radioactive series - the protactinium series'' \cite{1948Stu01}. Thorium metal targets were bombarded with 19~MeV deuterons and a 38~MeV $^4$He beam from the Berkeley 60-inch cyclotron forming $^{230}$Pa in (d,4n) and ($\alpha$,p5n) reactions. $^{222}$Ra was populated by subsequent $\alpha$ decay after the initial $\beta^-$ decay of $^{230}$Pa to $^{230}$U. Alpha-decay spectra were measured following chemical separation. ``It was necessary to measure the activity at such short intervals of time because of the short half-life of Ra$^{222}$, and as a result the statistical fluctuations were rather severe. An integral decay curve was plotted to minimize the scatter of points... The half-life of Ra$^{222}$ based on four such determinations is 38.0~seconds.'' The quoted half-life is included in the calculation of the currently adopted value of 38.0(5)~s.

\subsection*{$^{223}$Ra}
In the 1905 article ``A new radio-active product from actinium,'' Godlewski reported the discovery of a new activity in the natural actinium decay chain which was later identified as $^{223}$Ra \cite{1905God01}. The activity of actinium samples were measured following chemical separation. ``Taking into consideration the similarity of actinium and thorium, both as regards their chemical and radioactive properties, I resolved to try if the method used by Rutherford and Soddy for the separation of ThX would not serve also to separate an analogous product from actinium. The experiments were at once successful... This substance, which is so similar in properties to ThX, will be called actinium X (AcX). The product AcX, immediately after its separation, weight for weight, was more than a hundred times more active than the original actinium. The activity increased in the first day after removal to about 15 per cent.\ of its original value, and then decayed with the time according to an exponential law, falling to half value in about ten days.'' This half-life agrees with the currently accepted value of 11.43(5)~d. A year earlier Giesel had reported a new substance separated from emanium (actinium) \cite{1904Gie01} without any measurements \cite{1905God02}.

\subsection*{$^{224}$Ra}
In 1902 Rutherford and Soddy announced the discovery of a new activity extracted from thorium later identified as $^{224}$Ra in the paper ``The cause and nature of radioactivity - part I'' \cite{1902Rut01}. Activities from a thorium nitrate sample were observed with a photographic plate and an electrometer following chemical separation. ``If for present purposes the initial periods of the curves are disregarded and the later portions only considered, it will be seen at once that the time taken for the hydroxide to recover one half of its lost activity is about equal to the time taken by the ThX to lose half its activity, viz., in each case about 4~days, and speaking generally the percentage proportion of the lost activity regained by the hydroxide over any given interval is approximately equal to the percentage proportion of the activity lost by the ThX during the same interval.'' The quoted half-life is close to the currently adopted value of 3.66(4)~d.

\subsection*{$^{225}$Ra}
Hagemann et al.\ discovered $^{225}$Ra in 1947 in ``The (4n+1) radioactive series: the decay products of U$^{233}$'' \cite{1947Hag01}. The half-lives and $\alpha$- and $\beta$-decay energies of the nuclides in the decay chain of $^{233}$U were measured. ``These decay products, which constitute a substantial fraction of the entire missing, 4n+1, radioactive series are listed together with their radioactive properties, in [the table].'' The measured half-life of 14.8~d agrees with the presently accepted value of 14.9(2)~d. Hagemann et al.\ acknowledge the simultaneous observation of $^{225}$Ra by English et al.\ which was submitted only a day later and published in the same issue of Physical Review on the next page \cite{1947Eng01}.

\subsection*{$^{226}$Ra}
In 1898 Curie et al.\ announced the discovery of a new radioactive substance later identified as $^{226}$Ra in the paper ``Sur une nouvelle substance fortement radio-active, contenue dans la pechblende'' \cite{1898Cur02}. During the study of radioactivity of pitchblende in addition to polonium a second new radioactive element was chemically separated. ``La nouvelle substance radio-active que nous venons de trouver a toutes les apparences chimiques du baryum presque pur: elle n'est pr\'ecipit\'ee ni par l'hydrogene sulfure, ni par le sulfure d'ammonium, ni par l'ammoniaque; le sulfate est insoluble dans l'eau et dans les acides; le carbonate est insoluble dans l'eau; le chlorure, tr\`es soluble dans l'eau, est insoluble dans l'acide chlorhydrique concentr\'e et dans l'alcool. Enfin cette substance donne le spectre du baryum, facile \`a reconnaitre... Les diverses raisons que nous venons d'\'enumer\'er nous portent \`a croire que la nouvelle substance radio-active renferme un \'el\'ement nouveau, auquel nous proposons de donner le nom de radium.'' [The new radio-active substance which we have found has all the chemical appearances of almost pure barium: it is neither precipitated by hydrogen sulfide nor by the ammonium sulfide, nor with ammonia; the sulfate is insoluble in water and in acids; the carbonate is insoluble in water, the chloride, very soluble in water, is insoluble in concentrated hydrochloric acid and alcohol. Finally, the spectrum of this substance is easily recognizable as barium... The various reasons we have enumerated lead us to believe that the new radioactive substance contains a new element, which we propose to name radium.] The currently adopted half-life is 1600(7)~y.

\subsection*{$^{227}$Ra}
$^{227}$Ra was discovered by Butler and Adam and the results were published in the 1953 article ``Radiations of Ra$^{227}$'' \cite{1953But01}. $^{226}$Ra was irradiated with thermal neutrons from the Chalk River NRX reactor producing $^{227}$Ra in neutron capture reactions. Beta- and gamma-rays were measured with an end-window Geiger counter and a sodium iodide (thalliated) scintillation spectrometer, respectively, following chemical separation. ``Ra$^{227}$ is a $\beta^-$ emitter with a half-life of 41.2$\pm$0.2~minutes and with a $\beta^-$ endpoint of 1.31$\pm$0.02~Mev.'' This half-life agrees with the currently accepted value of 42.2(5)~min.

\subsection*{$^{228}$Ra}
Hahn first observed a new activity in the natural thorium decay series later identified as $^{228}$Ra in 1907 and published his results in the article ``Ein neues Zwischenprodukt im Thorium'' \cite{1907Hah01}. The activities from thorium nitrate samples of different ages were compared. ``Es ist naheliegend, da\ss\ die allm\"ahliche Abnahme der Aktivit\"at der frisch bereiteten Thorpr\"aparate vom Zerfall eines aus irgend einem Grunde vorliegenden \"Uberschusses an Radiothor herr\"uhrt; und in der Tat ergibt sich als Gr\"o\ss enordnung f\"ur die Abnahme eine Periode von etwa zwei Jahren... Von der definitiven Wahl eines Namens f\"ur den neuen K\"orper sehe ich ab, bis sich seine Natur genauer hat feststellen lassen. Im letzteren Fall w\"urde mir der Name `Mesothorium' als zweckm\"a\ss ig erscheinen.'' [It seems evident that the slow reduction of the activity of the newly prepared thorium samples is due to the decay of an unexplained excess of radio thorium; and indeed the order of magnitude of the activity is about two years... I restrain from selecting a name for the new substance until its properties are better determined. In the latter case the name `mesothorium' seems suitable.] Hahn submitted the identical paper three days later to another journal \cite{1907Hah02}. In a subsequent paper Hahn estimated the half-life of mesothorium to be about 7~years \cite{1907Hah03}. A year later Hahn quoted a half-life of $\sim$5.5~y \cite{1908Hah01} which agrees with the presently accepted value of 5.75(3)~y for $^{228}$Ra. In the same paper Hahn renamed mesothorium to mesothorium 1 because he identified a separated decay product (mesothorium 2 or $^{228}$Ac).

\subsection*{$^{229}$Ra}
In 1975 the discovery of $^{229}$Fr by Ravn et al.\ was announced in the paper ``Short-lived isotopes of alkali and alkaline-earth elements studied by on-line isotope separator techniques'' \cite{1975Rav01}. A thorium plus lanthanum target was bombarded with protons from the CERN synchrocyclotron. Beta-ray decay curves were measured with a 4$\pi$ plastic detector following mass separation with the isotope separator ISOLDE. ``The following half-lives of new nucleides have been determined: ... $^{229}$Ra (4.0$\pm$0.2)~min.'' This half-life is the current value.

% Explore the 1935Hah01 paper who assigned a 1min half-life to 229Ra and lost of subsequent papers...

\subsection*{$^{230}$Ra}
In the 1978 article ``Decay of $^{230}$Ra and $^{230}$Ac,'' Gilat and Katcoff announced the discovery of $^{230}$Ra \cite{1978Gil01}. At the Brookhaven Medium Energy Intense Neutron (MEIN) facility thorium targets were irradiated with secondary neutrons produced by stopping 200 MeV protons from the Brookhaven AGS injector Linac in a copper beam stop. $^{230}$Ra was produced in $^{232}$Th(n,2pn) reactions. Gamma-ray spectra and decay curves were measured following chemical separation. ``The half-life of $^{230}$Ra was determined by: (1) analysis of the decay curves of $\sim$20 of the most prominent lines in the $^{230}$Ra-$^{230}$Ac spectrum, some of which are shown in [the figure]; and (2) milking experiments, in which the $^{23°}$Ac daughter was milked successively at 20~min intervals over a period of 3~hr, and the parent half-life was extracted from the values of the daughter activity. Both methods gave the identical result of 93$\pm$2~min.'' This half-life is the currently adopted value. The tentative assignment of a 45.5(15)~min half-life to $^{230}$Ra was evidently incorrect \cite{1975Rav01}. An even earlier measurement of a $\sim$1~hr half-life assigned to $^{230}$Ra was only published in a conference abstract \cite{1952Jen01}.

\subsection*{$^{231,232}$Ra}
The observation of $^{231}$Ra and $^{232}$Ra was reported by Ahmad et al.\ in the 1983 paper ``Determination of nuclear spins and moments in a series of radium isotopes'' \cite{1983Ahm01}. A UC$_2$ target was bombarded with 600~MeV protons from the CERN synchrocyclotron. $^{231}$Ra and $^{232}$Ra were identified by on-line collinear fast beam laser spectroscopy following mass separation with the isotope separator ISOLDE. ``The optical high-resolution measurements on 18 Ra isotopes in the range A = 208--232 have been carried out in the 4826~\AA~line (7s$^2$ $^1$S$_0$-7s7p$^1$P$_1$) of the neutral Ra and the 4683~\AA~line (7s $^2$S$_{1/2}$-7p$^{2}$P$_{1/2}$) of the Ra$^+$ spectra.'' The presently accepted half-lives for $^{213}$Ra and $^{232}$Ra are 103(3)~s and 250(50)~s.

\subsection*{$^{233,234}$Ra}
Mezlev et al.\ reported the discovery of $^{233}$Ra and $^{234}$Ra in the 1990 paper ``Search for delayed fission in neutron-rich nuclei'' \cite{1990Mez01}. A uranium target was bombarded with 1~GeV protons. Beta-, gamma-, and X-rays were measured with solid state detectors following mass separation with the on-line mass separator IRIS. ``Due to this technique the new isotopes $^{232}$Fr (T$_{1/2}$=5$\pm$1~s), $^{233}$Ra(T$_{1/2}$=30$\pm$5~s) and $^{234}$Ra (T$_{1/2}$=30$\pm$10~s) have been identified using the solid state detectors for the registration of the beta-, gamma-, X-radiation.'' For $^{233}$Ra, the quoted half-life is the currently accepted value, and for $^{234}$Ra the quoted half-life agrees with the currently adopted value of 30(10)~s.

%*********************************************************
%*********************************************************

\section{Summary}
The discoveries of the known astatine, radon, francium, and radium isotopes have been compiled and the methods of their production discussed. Only the following six isotopes were initially identified incorrectly: $^{204,206,216}$At, $^{214}$Rn, and $^{202,230}$Ra. In addition, the half-lives of $^{200,201,211}$At and $^{205,212}$Ra were at first reported without definite mass assignments.

\ack

This work was supported by the National Science Foundation under grants No. PHY06-06007 (NSCL) and PHY10-62410 (REU).

%%% Here we use thebibliography environment to produce the reference list,
%%% but you can use BibTeX as well:
\bibliography{../isotope-discovery-references}

\newpage

%%% Please start a new page by uncommenting the next
\newpage

\TableExplanation

\bigskip
\renewcommand{\arraystretch}{1.0}

\section{Table 1.\label{tbl1te} Discovery of astatine, radon, francium, and radium isotopes }
\begin{tabular*}{0.95\textwidth}{@{}@{\extracolsep{\fill}}lp{5.5in}@{}}
\multicolumn{2}{p{0.95\textwidth}}{ }\\

Isotope & Astatine, radon, francium, and radium isotope \\
First author & First author of refereed publication \\
Journal & Journal of publication \\
Ref. & Reference \\
Method & Production method used in the discovery: \\

  & FE: fusion evaporation \\
  & NC: Neutron capture reactions \\
  & LP: light-particle reactions (including neutrons) \\
  & RD: radioactive decay \\
  & SP: spallation reactions \\
  & PF: projectile fragmentation of fission \\

Laboratory & Laboratory where the experiment was performed\\
Country & Country of laboratory\\
Year & Year of discovery \\
\end{tabular*}
\label{tableI}

\datatables % This command is necessary to get the table names in toc

%% One-page data tables are also best formatted using the longtable
%% environment:
%\begin{longtable}{c}
%\caption{This is the First Data Table}\\
%\endhead\\
%\end{longtable}

%% If the table is to span over the whole text width, we set:

\setlength{\LTleft}{0pt}
\setlength{\LTright}{0pt}

% To avoid ``Overfull \hboxes...'' decrease the intercolumn spacing:

\setlength{\tabcolsep}{0.5\tabcolsep}

\renewcommand{\arraystretch}{1.0}

\footnotesize % we need to squeeze the font size a lot!

\begin{longtable}{@{\extracolsep\fill}llllllll@{}}
\caption{Discovery of astatine, radon, francium, and radium Isotopes. See page\ \pageref{tbl1te} for Explanation of Tables}
Isotope & First Author & Journal & Ref. & Method & Laboratory & Country & Year\\
\hline\\
\endfirsthead\\
\caption[]{(continued)}
Isotope & First author & Journal & Ref. & Method & Laboratory & Country & Year\\
\hline\\
\endhead
$^{191}$At & H. Kettunen & Eur. Phys. J. A &\cite{2003Ket01}& FE & Jyv\"askyl\"a & Finland &2003 \\
$^{192}$At & A.N. Andreyev & Phys. Rev. C &\cite{2006And01}& FE & Darmstadt & Germany &2006 \\
$^{193}$At & H. Kettunen & Eur. Phys. J. A &\cite{2003Ket01}& FE & Jyv\"askyl\"a & Finland &2003 \\
$^{194}$At & A.N. Andreyev & Phys. Rev. C &\cite{2009And01}& FE & Darmstadt & Germany &2009 \\
$^{195}$At & Y. Tagaya & Eur. Phys. J. A &\cite{1999Tag01}& FE & RIKEN & Japan &1999 \\
$^{196}$At & W. Treytl & Nucl. Phys. A &\cite{1967Tre01}& FE & Berkeley & USA &1967 \\
$^{197}$At & W. Treytl & Nucl. Phys. A &\cite{1967Tre01}& FE & Berkeley & USA &1967 \\
$^{198}$At & W. Treytl & Nucl. Phys. A &\cite{1967Tre01}& FE & Berkeley & USA &1967 \\
$^{199}$At & W. Treytl & Nucl. Phys. A &\cite{1967Tre01}& FE & Berkeley & USA &1967 \\
$^{200}$At & R.W. Hoff & J. Inorg. Nucl. Chem. &\cite{1963Hof01}& FE & Berkeley & USA &1963 \\
$^{201}$At & R.W. Hoff & J. Inorg. Nucl. Chem. &\cite{1963Hof01}& FE & Berkeley & USA &1963 \\
$^{202}$At & R.M. Latimer & J. Inorg. Nucl. Chem. &\cite{1961Lat01}& FE & Berkeley & USA &1961 \\
$^{203}$At & G.W. Barton & Phys. Rev. &\cite{1951Bar01}& LP & Berkeley & USA &1951 \\
$^{204}$At & R.M. Latimer & J. Inorg. Nucl. Chem. &\cite{1961Lat01}& FE & Berkeley & USA &1961 \\
$^{205}$At & G.W. Barton & Phys. Rev. &\cite{1951Bar01}& LP & Berkeley & USA &1951 \\
$^{206}$At & R.M. Latimer & J. Inorg. Nucl. Chem. &\cite{1961Lat01}& FE & Berkeley & USA &1961 \\
$^{207}$At & G.W. Barton & Phys. Rev. &\cite{1951Bar01}& LP & Berkeley & USA &1951 \\
$^{208}$At & E.K. Hyde & Phys. Rev. &\cite{1950Hyd01}& LP & Berkeley & USA &1950 \\
$^{209}$At & G.W. Barton & Phys. Rev. &\cite{1951Bar01}& LP & Berkeley & USA &1951 \\
$^{210}$At & E.L. Kelly & Phys. Rev. &\cite{1949Kel01}& LP & Berkeley & USA &1949 \\
$^{211}$At & D.R. Corson & Phys. Rev. &\cite{1940Cor01}& LP & Berkeley & USA &1940 \\
$^{212}$At & M.M. Winn & Proc. Roy. Soc. A &\cite{1954Win01}& LP & Birmingham & UK &1954 \\
$^{213}$At & R.L. Hahn & Nucl. Phys. A &\cite{1968Hah01}& LP & Oak Ridge & USA &1968 \\
$^{214}$At & W.W. Meinke & Phys. Rev. &\cite{1949Mei01}& LP & Berkeley & USA &1949 \\
$^{215}$At & B. Karlik & Naturwiss. &\cite{1944Kar01}& RD & Wien & Austria &1944 \\
$^{216}$At & A. Ghiorso & Phys. Rev. &\cite{1948Ghi01}& LP & Berkeley & USA &1948 \\
$^{217}$At & F. Hagemann & Phys. Rev. &\cite{1947Hag01}& RD & Argonne & USA &1947 \\
$^{218}$At & B. Karlik & Naturwiss. &\cite{1943Kar01}& RD & Wien & Austria &1943 \\
$^{219}$At & E.K. Hyde & Phys. Rev. &\cite{1953Hyd01}& RD & Berkeley & USA &1953 \\
$^{220}$At & C.F. Liang & J. Phys. G &\cite{1989Lia01}& SP & Orsay & France &1989 \\
$^{221}$At & D.G. Burke & Z. Phys. A &\cite{1989Bur01}& SP & CERN & Switzerland &1989 \\
$^{222}$At & D.G. Burke & Z. Phys. A &\cite{1989Bur01}& SP & CERN & Switzerland &1989 \\
$^{223}$At & D.G. Burke & Z. Phys. A &\cite{1989Bur01}& SP & CERN & Switzerland &1989 \\
$^{224}$At & L. Chen & Phys. Lett. B &\cite{2010Che01}& PF & Darmstadt & Germany &2010 \\
$^{225}$At & H. Alvarez-Pol & Phys. Rev. C &\cite{2010Alv01}& PF & Darmstadt & Germany &2010 \\
$^{226}$At & H. Alvarez-Pol & Phys. Rev. C &\cite{2010Alv01}& PF & Darmstadt & Germany &2010 \\
$^{227}$At & H. Alvarez-Pol & Phys. Rev. C &\cite{2010Alv01}& PF & Darmstadt & Germany &2010 \\
$^{228}$At & H. Alvarez-Pol & Phys. Rev. C &\cite{2010Alv01}& PF & Darmstadt & Germany &2010 \\
$^{229}$At & H. Alvarez-Pol & Phys. Rev. C &\cite{2010Alv01}& PF & Darmstadt & Germany &2010 \\
 & & & & & &  \\
 & & & & & &  \\
$^{193}$Rn & A.N. Andreyev & Phys. Rev. C &\cite{2006And02}& FE & Darmstadt & Germany &2006 \\
$^{194}$Rn & A.N. Andreyev & Phys. Rev. C &\cite{2006And02}& FE & Darmstadt & Germany &2006 \\
$^{195}$Rn & H. Kettunen & Phys. Rev. C &\cite{2001Ket01}& FE & Jyv\"askyl\"a & Finland &2001 \\
$^{196}$Rn & K. Morita & Z. Phys. A &\cite{1995Mor01}& FE & RIKEN & Japan &1995 \\
$^{197}$Rn & K. Morita & Z. Phys. A &\cite{1995Mor01}& FE & RIKEN & Japan &1995 \\
$^{198}$Rn & F. Calaprice & Phys. Rev. C &\cite{1984Cal01}& SP & CERN & Switzerland &1984 \\
$^{199}$Rn & A.C. DiRienzo & Phys. Rev. C &\cite{1980DiR01}& FE & Brookhaven & USA &1980 \\
$^{200}$Rn & P. Hornshoj & Nucl. Phys. A &\cite{1971Hor01}& SP & CERN & Switzerland &1971 \\
$^{201}$Rn & K. Valli & Phys. Rev. &\cite{1967Val01}& FE & Berkeley & USA &1967 \\
$^{202}$Rn & K. Valli & Phys. Rev. &\cite{1967Val01}& FE & Berkeley & USA &1967 \\
$^{203}$Rn & K. Valli & Phys. Rev. &\cite{1967Val01}& FE & Berkeley & USA &1967 \\
$^{204}$Rn & K. Valli & Phys. Rev. &\cite{1967Val01}& FE & Berkeley & USA &1967 \\
$^{205}$Rn & K. Valli & Phys. Rev. &\cite{1967Val01}& FE & Berkeley & USA &1967 \\
$^{206}$Rn & W.E. Burcham & Proc. Phys. Soc. A &\cite{1954Bur01}& FE & Birmingham & UK &1954 \\
$^{207}$Rn & W.E. Burcham & Proc. Phys. Soc. A &\cite{1954Bur01}& FE & Birmingham & UK &1954 \\
$^{208}$Rn & F.F. Momyer& J. Inorg. Nucl. Chem. &\cite{1955Mom01}& SP & Berkeley & USA &1955 \\
$^{209}$Rn & F.F. Momyer& Phys. Rev. &\cite{1952Mom01}& SP & Berkeley & USA &1952 \\
$^{210}$Rn & F.F. Momyer& Phys. Rev. &\cite{1952Mom01}& SP & Berkeley & USA &1952 \\
$^{211}$Rn & F.F. Momyer& Phys. Rev. &\cite{1952Mom01}& SP & Berkeley & USA &1952 \\
$^{212}$Rn & E.K. Hyde & Phys. Rev. &\cite{1950Hyd01}& SP & Berkeley & USA &1950 \\
$^{213}$Rn & H. Rotter & Sov. J. Nucl. Phys. &\cite{1967Rot01}& FE & Dubna & Russia &1967 \\
$^{214}$Rn & D.F. Torgerson & Nucl. Phys. A &\cite{1970Tor01}& FE & Yale & USA &1970 \\
$^{215}$Rn & W.W. Meinke & Phys. Rev. &\cite{1952Mei01}& LP & Berkeley & USA &1952 \\
$^{216}$Rn & W.W. Meinke & Phys. Rev. &\cite{1949Mei01}& LP & Berkeley & USA &1949 \\
$^{217}$Rn & W.W. Meinke & Phys. Rev. &\cite{1949Mei01}& LP & Berkeley & USA &1949 \\
$^{218}$Rn & M.H. Studier & Phys. Rev. &\cite{1948Stu01}& LP & Argonne & USA &1948 \\
$^{219}$Rn & F. Giesel & Ber. Deuts. Chem. Ges. &\cite{1903Gie01}& RD & Braunschweig & Germany &1903 \\
$^{220}$Rn & E. Rutherford & Phil. Mag. &\cite{1900Rut01}& RD & McGill & Canada &1900 \\
$^{221}$Rn & F.F. Momyer& Phys. Rev. &\cite{1956Mom01}& LP & Berkeley & USA &1956 \\
$^{222}$Rn & P. Curie & Compt. Rend. Acad. Sci. &\cite{1899Cur01}& RD & Paris & France &1899 \\
$^{223}$Rn & F.D.S. Butement & J. Inorg. Nucl. Chem. &\cite{1964But01}& LP & Liverpool & UK &1964 \\
$^{224}$Rn & F.D.S. Butement & J. Inorg. Nucl. Chem. &\cite{1964But01}& LP & Liverpool & UK &1964 \\
$^{225}$Rn & P.G. Hansen & Phys. Lett. B &\cite{1969Han01}& SP & CERN & Switzerland &1969 \\
$^{226}$Rn & P.G. Hansen & Phys. Lett. B &\cite{1969Han01}& SP & CERN & Switzerland &1969 \\
$^{227}$Rn & M.J.G. Borge & Z. Phys. A &\cite{1986Bor01}& SP & CERN & Switzerland &1986 \\
$^{228}$Rn & M.J.G. Borge & Z. Phys. A &\cite{1989Bor01}& SP & CERN & Switzerland &1989 \\
$^{229}$Rn & D. Neidherr & Phys. Rev. Lett. &\cite{2009Nei01}& SP & CERN & Switzerland &2009 \\
$^{230}$Rn & H. Alvarez-Pol & Phys. Rev. C &\cite{2010Alv01}& PF & Darmstadt & Germany &2010 \\
$^{231}$Rn & H. Alvarez-Pol & Phys. Rev. C &\cite{2010Alv01}& PF & Darmstadt & Germany &2010 \\
 & & & & & &  \\
 & & & & & &  \\
$^{199}$Fr & Y. Tagaya & Eur. Phys. J. A &\cite{1999Tag01}& FE & RIKEN & Japan &1999 \\
$^{200}$Fr & K. Morita & Z. Phys. A &\cite{1995Mor01}& FE & RIKEN & Japan &1995 \\
$^{201}$Fr & G.T. Ewan & Z. Phys. A &\cite{1980Ewa01}& SP & CERN & Switzerland &1980 \\
$^{202}$Fr & G.T. Ewan & Z. Phys. A &\cite{1980Ewa01}& SP & CERN & Switzerland &1980 \\
$^{203}$Fr & K. Valli & J. Inorg. Nucl. Chem. &\cite{1967Val02}& FE & Berkeley & USA &1967 \\
$^{204}$Fr & R.D. Griffioen & Phys. Rev. &\cite{1964Gri01}& FE & Berkeley & USA &1964 \\
$^{205}$Fr & R.D. Griffioen & Phys. Rev. &\cite{1964Gri01}& FE & Berkeley & USA &1964 \\
$^{206}$Fr & R.D. Griffioen & Phys. Rev. &\cite{1964Gri01}& FE & Berkeley & USA &1964 \\
$^{207}$Fr & R.D. Griffioen & Phys. Rev. &\cite{1964Gri01}& FE & Berkeley & USA &1964 \\
$^{208}$Fr & R.D. Griffioen & Phys. Rev. &\cite{1964Gri01}& FE & Berkeley & USA &1964 \\
$^{209}$Fr & R.D. Griffioen & Phys. Rev. &\cite{1964Gri01}& FE & Berkeley & USA &1964 \\
$^{210}$Fr & R.D. Griffioen & Phys. Rev. &\cite{1964Gri01}& FE & Berkeley & USA &1964 \\
$^{211}$Fr & R.D. Griffioen & Phys. Rev. &\cite{1964Gri01}& FE & Berkeley & USA &1964 \\
$^{212}$Fr & E.K. Hyde & Phys. Rev. &\cite{1950Hyd01}& SP & Berkeley & USA &1950 \\
$^{213}$Fr & R.D. Griffioen & Phys. Rev. &\cite{1964Gri01}& FE & Berkeley & USA &1964 \\
$^{214}$Fr & H. Rotter & Sov. J. Nucl. Phys. &\cite{1967Rot01}& FE & Dubna & Russia &1967 \\
$^{215}$Fr & J. Borggreen & Phys. Rev. C &\cite{1970Bor01}& FE & Berkeley & USA &1970 \\
$^{216}$Fr & J. Borggreen & Phys. Rev. C &\cite{1970Bor01}& FE & Berkeley & USA &1970 \\
$^{217}$Fr & R.L. Hahn & Nucl. Phys. A &\cite{1968Hah01}& LP & Oak Ridge & USA &1968 \\
$^{218}$Fr & W.W. Meinke & Phys. Rev. &\cite{1949Mei01}& LP & Berkeley & USA &1949 \\
$^{219}$Fr & A. Ghiorso & Phys. Rev. &\cite{1948Ghi01}& LP & Berkeley & USA &1948 \\
$^{220}$Fr & A. Ghiorso & Phys. Rev. &\cite{1948Ghi01}& LP & Berkeley & USA &1948 \\
$^{221}$Fr & F. Hagemann & Phys. Rev. &\cite{1947Hag01}& RD & Argonne & USA &1947 \\
$^{222}$Fr & L. Westgaard & Z. Phys. A &\cite{1975Wes01}& SP & CERN & Switzerland &1975 \\
$^{223}$Fr & M. Perey & Compt. Rend. Acad. Sci. &\cite{1939Per01}& RD & Paris & France &1939 \\
$^{224}$Fr & P.G. Hansen & Phys. Lett. B &\cite{1969Han01}& SP & CERN & Switzerland &1969 \\
$^{225}$Fr & P.G. Hansen & Phys. Lett. B &\cite{1969Han01}& SP & CERN & Switzerland &1969 \\
$^{226}$Fr & P.G. Hansen & Phys. Lett. B &\cite{1969Han01}& SP & CERN & Switzerland &1969 \\
$^{227}$Fr & R. Klapisch & Phys. Rev. Lett. &\cite{1972Kla01}& SP & CERN & Switzerland &1972 \\
$^{228}$Fr & R. Klapisch & Phys. Rev. Lett. &\cite{1972Kla01}& SP & CERN & Switzerland &1972 \\
$^{229}$Fr & H.L. Ravn & J. Inorg. Nucl. Chem. &\cite{1975Rav01}& SP & CERN & Switzerland &1975 \\
$^{230}$Fr & W. Kurcewicz & Nucl. Phys. A &\cite{1987Kur01}& SP & CERN & Switzerland &1987 \\
$^{231}$Fr & P. Hill & Z. Phys. A &\cite{1985Hil01}& SP & CERN & Switzerland &1985 \\
$^{232}$Fr & K.A. Mezlev & Z. Phys. A &\cite{1990Mez01}& SP & Leningrad & Russia &1990 \\
$^{233}$Fr & H. Alvarez-Pol & Phys. Rev. C &\cite{2010Alv01}& PF & Darmstadt & Germany &2010 \\
 & & & & & &  \\
 & & & & & &  \\
$^{201}$Ra & J. Uusitalo & Phys. Rev. C &\cite{2005Uus01}& FE & Jyv\"askyl\"a & Finland &2005 \\
$^{202}$Ra & J. Uusitalo & Phys. Rev. C &\cite{2005Uus01}& FE & Jyv\"askyl\"a & Finland &2005 \\
$^{203}$Ra & M. Leino & Z. Phys. A &\cite{1996Lei01}& FE & Jyv\"askyl\"a & Finland &1996 \\
$^{204}$Ra & M.J. Leddy & Phys. Rev. C &\cite{1995Led01}& FE & Argonne & USA &1995 \\
$^{205}$Ra & F.P. He\ss berger & Europhys. Lett. &\cite{1987Hes01}& FE & Darmstadt & Germany &1987 \\
$^{206}$Ra & K. Valli & Phys. Rev. &\cite{1967Val03}& FE & Berkeley & USA &1967 \\
$^{207}$Ra & K. Valli & Phys. Rev. &\cite{1967Val03}& FE & Berkeley & USA &1967 \\
$^{208}$Ra & K. Valli & Phys. Rev. &\cite{1967Val03}& FE & Berkeley & USA &1967 \\
$^{209}$Ra & K. Valli & Phys. Rev. &\cite{1967Val03}& FE & Berkeley & USA &1967 \\
$^{210}$Ra & K. Valli & Phys. Rev. &\cite{1967Val03}& FE & Berkeley & USA &1967 \\
$^{211}$Ra & K. Valli & Phys. Rev. &\cite{1967Val03}& FE & Berkeley & USA &1967 \\
$^{212}$Ra & K. Valli & Phys. Rev. &\cite{1967Val03}& FE & Berkeley & USA &1967 \\
$^{213}$Ra & F.F. Momyer& J. Inorg. Nucl. Chem. &\cite{1955Mom01}& SP & Berkeley & USA &1955 \\
$^{214}$Ra & H. Rotter & Sov. J. Nucl. Phys. &\cite{1967Rot01}& FE & Dubna & Russia &1967 \\
$^{215}$Ra & H. Rotter & Sov. J. Nucl. Phys. &\cite{1967Rot01}& FE & Dubna & Russia &1967 \\
$^{216}$Ra & T. Nomura & Phys. Lett. B &\cite{1972Nom01}& FE & RIKEN & Japan &1972 \\
$^{217}$Ra & D.F. Torgerson & Nucl. Phys. A &\cite{1970Tor01}& FE & Yale & USA &1970 \\
$^{218}$Ra & D.F. Torgerson & Nucl. Phys. A &\cite{1970Tor01}& FE & Yale & USA &1970 \\
$^{219}$Ra & W.W. Meinke & Phys. Rev. &\cite{1952Mei01}& LP & Berkeley & USA &1952 \\
$^{220}$Ra & W.W. Meinke & Phys. Rev. &\cite{1949Mei01}& LP & Berkeley & USA &1949 \\
$^{221}$Ra & W.W. Meinke & Phys. Rev. &\cite{1949Mei01}& LP & Berkeley & USA &1949 \\
$^{222}$Ra & M.H. Studier & Phys. Rev. &\cite{1948Stu01}& LP & Argonne & USA &1948 \\
$^{223}$Ra & T. Godlewski & Nature &\cite{1905God01}& RD & McGill & Canada &1905 \\
$^{224}$Ra & E. Rutherford & Phil. Mag. &\cite{1902Rut01}& RD & McGill & Canada &1902 \\
$^{225}$Ra & F. Hagemann & Phys. Rev. &\cite{1947Hag01}& RD & Argonne & USA &1947 \\
$^{226}$Ra & P. Curie & Compt. Rend. Acad. Sci. &\cite{1898Cur02}& RD & Paris & France &1898 \\
$^{227}$Ra & J.P. Butler & Phys. Rev. &\cite{1953But01}& NC & Chalk River & Canada &1953 \\
$^{228}$Ra & O. Hahn & Ber. Deuts. Chem. Ges. &\cite{1907Hah01}& RD & Berlin & Germany &1907 \\
$^{229}$Ra & H.L. Ravn & J. Inorg. Nucl. Chem. &\cite{1975Rav01}& SP & CERN & Switzerland &1975 \\
$^{230}$Ra & J. Gilat & J. Inorg. Nucl. Chem. &\cite{1978Gil01}& LP & Brookhaven & USA &1978 \\
$^{231}$Ra & S.A. Ahmad & Phys. Lett. B &\cite{1983Ahm01}& SP & CERN & Switzerland &1983 \\
$^{232}$Ra & S.A. Ahmad & Phys. Lett. B &\cite{1983Ahm01}& SP & CERN & Switzerland &1983 \\
$^{233}$Ra & K.A. Mezlev & Z. Phys. A &\cite{1990Mez01}& SP & Leningrad & Russia &1990 \\
$^{234}$Ra & K.A. Mezlev & Z. Phys. A &\cite{1990Mez01}& SP & Leningrad & Russia &1990 \\
\end{longtable}

\end{document}